\documentclass[%
 reprint,
superscriptaddress,
nofootinbib,
 amsmath,amssymb,
 onecolumn,
prx,
]{revtex4-2}

\usepackage{graphicx}
\usepackage{dcolumn}
\usepackage{bm}
\usepackage[colorlinks]{hyperref}

\usepackage{cmbright}
\DeclareFontShape{OT1}{cmss}{m}{it}{<->ssub*cmss/m/sl}{}

\usepackage{color}
\usepackage[caption=false]{subfig}

\renewcommand{\vec}[1]{\boldsymbol{#1}}

\newcommand{\Mpch}{h^{-1}\mathrm{Mpc}}
\newcommand{\hMpc}{h\,\mathrm{Mpc}^{-1}}

\newcommand{\vk}{\vec k}

\newcommand{\vx}{\vec x}
\newcommand{\rmd}{\mathrm{d}}

\newcommand{\vy}{\vec y}
\renewcommand{\vr}{\vec r}

\newcommand{\hr}{\hat{\vec r}}
\newcommand{\hn}{\hat{\vec n}}

\usepackage{color}
\def\beq{\begin{eqnarray}}
\def\eeq{\end{eqnarray}}

\definecolor{darkgreen}{RGB}{0,120,0}
\definecolor{darkred}{RGB}{150,30,30}
\newcommand{\new}[1]{#1}
\newcommand{\resub}[1]{#1}
\newcommand{\resubtwo}[1]{#1}
\usepackage{fixltx2e}
\MakeRobust{\newline}
\def\mnras{\rm{MNRAS}}

\def\aap{\rm{A\&A}}
\def\baas{\rm{BAAS}}

\def\apj{\rm{ApJ}}                 
\def\apjl{\rm{ApJ}}                
\def\jcap{\rm{JCAP}}

\begin{document}


\title{%
The Disordered Heterogeneous Universe: \texorpdfstring{\\}{}
Galaxy Distribution and Clustering Across Length Scales
}

\author{Oliver H.\,E. Philcox}
\email{ohep2@cantab.ac.uk}
\affiliation{Department of Astrophysical Sciences, Princeton University, Princeton, NJ 08540, USA}%
\affiliation{School of Natural Sciences, Institute for Advanced Study, 1 Einstein Drive, Princeton, NJ 08540, USA}
\affiliation{Center for Theoretical Physics, Department of Physics,
Columbia University, New York, NY 10027, USA}
\affiliation{Simons Society of Fellows, Simons Foundation, New York, NY 10010, USA}
\author{Salvatore Torquato}
\email{torquato@electron.princeton.edu}
\affiliation{School of Natural Sciences, Institute for Advanced Study, 1 Einstein Drive, Princeton, NJ 08540, USA}
\affiliation{Department of Chemistry, Department of Physics, Princeton Institute for the Science and Technology of Materials, and Program in Applied and Computational Mathematics, Princeton University, Princeton, New Jersey 08540, USA}%


\begin{abstract}
The studies of disordered heterogeneous media and galaxy cosmology share a common goal: analyzing the disordered distribution of particles and/or building blocks at `microscales' to predict physical properties of the medium at `macroscales', whether it be a liquid, colloidal suspension, composite material, galaxy cluster, or entire Universe. The theory of disordered heterogeneous media provides an array of theoretical and computational techniques to characterize a wide class of complex material microstructures. In this work, we apply them to describe the disordered distributions of galaxies \resubtwo{obtained from recent suites of dark matter simulations}. We focus on the determination of lower-order correlation functions, `void' and `particle' nearest-neighbor functions, certain cluster statistics, pair-connectedness functions, percolation properties, and a scalar order metric to quantify the degree of order. Compared to analogous homogeneous Poisson and typical disordered systems, the cosmological simulations exhibit enhanced large-scale clustering and longer tails in the void and particle nearest-neighbor functions, due to the presence of quasi-long-range correlations imprinted by early Universe physics, with a minimum particle separation far below the mean nearest-neighbor distance. On large scales, the system appears `hyperuniform', as a result of primordial density fluctuations, whilst on the smallest scales, the system becomes almost `antihyperuniform', as evidenced by its number variance.  Additionally, via a finite scaling analysis, we compute the percolation threshold of the galaxy catalogs, finding this to be significantly lower than for Poisson realizations (at reduced density $\eta_c = 0.25$ in our fiducial analysis compared to $\eta_c = 0.34$), with strong dependence on the mean density; this is consistent with the observation that the galaxy distribution contains voids of up to $50\%$ larger radius. However, the two sets of simulations appear to share the same fractal dimension on scales much larger than the average inter-galaxy separation, implying that they lie in the same universality class. We also show that the distribution of galaxies are a highly correlated disordered system (relative to the uncorrelated Poisson distribution), as measured by the $\tau$ order metric. Finally, we consider the ability of large-scale clustering statistics to constrain cosmological parameters, such as the Universe's expansion rate, using simulation-based inference. Both the nearest-neighbor distribution and pair-connectedness function (which includes contributions from correlation functions of all order) are found to considerably tighten bounds on the amplitude of quantum-mechanical fluctuations from inflation at a level equivalent to observing twenty-five times more galaxies. \resubtwo{The pair-connectedness function in particular} provides a useful alternative to the standard three-particle correlation, \resubtwo{since it contains similar large-scale information to the three-point function, can be computed highly efficiently, and straightforwardly extended to small scales} (though likely requires simulation-based modeling). This work provides the first application of such techniques to cosmology, providing both a novel system to test heterogeneous media descriptors, and a tranche of new tools for cosmological analyses. \new{A range of extensions are possible, including implementation on observational data; this will require further study on \resubtwo{various observational effects, necessitating high-resolution simulations.}
}
\end{abstract}

\maketitle


\section{Introduction}\label{sec: intro}





From condensed phases of matter to ecological systems to the primordial distribution of matter in the universe, Nature abounds with examples of disordered arrangements of interacting entitites that form structures
with diverse geometries and topologies. To understand the collective behavior of such phenomena, it is vital to have a mathematical formalism that enables a stochastic description of the constituent objects, particularly with regards to their spatial distribution and clustering, whether they be carbon atoms, concrete conglomerates or individual galaxies. The theory of disordered heterogeneous media \cite{To02a,Sa03}, which
includes techniques from statistical mechanics \cite{Han13},  provides a natural and powerful machinery with which to equip ourselves in this venture. In particular, its
primary objective is to connect the properties of the interacting constituents to their large-scale attributes, such as a material's bulk transport,
mechanical and electromagnetic properties. This is rigorously done by generally relating the bulk properties  to an infinite set of diverse types of  statistical correlation functions that characterize the microstructures \cite{To02a}, including those that contain topological information, such as phase connectivity and percolation characteristics. While
this theoretical machinery has been primarily applied to earth-bound materials, their applicability is far from terrestrial: these techniques work similarly for any phenomenon that can be treated as complex disordered heterogeneous media \cite{To02a}, including the spatial distribution and clustering of galaxies. 

Whilst the bulk of cosmological research in the past two decades has focused on analysis of the `cosmic microwave background' (the radiation signature of physics in the first $\sim 300\,000$ years, which provides a snapshot of the early Universe), that concerning the distribution of galaxies using statistical descriptors has become progressively more important \cite{peebles80,Sa00,Ga05}, particularly with the advent of large-scale surveys, such as the forthcoming Dark Energy Spectroscopic Instrument (DESI) \citep{desi16} and the Euclid satellite \citep{euclid11}. The distribution of galaxies traces the distribution of matter in the early Universe \citep[e.g.,][]{peebles80}; as such, it encodes information on a wealth of cosmological parameters, such as the density of matter. An open question is how best to analyze the data: most works focus on measuring the \textit{correlation functions} of the galaxy distribution, and comparing them to physical models \resub{(either explicitly derived, or numerically simulated) \citep[e.g.,][]{Bernardeau:2001qr}}, though this is known to be suboptimal in terms of information content. Whilst a number of alternative statistics have been proposed (including void statistics \citep[e.g.,][]{Sheth:2003py,Pisani:2019cvo}, marked density fields \citep[e.g.,][]{,2005MNRAS.364..796S,2016JCAP...11..057W,2021PhRvL.126a1301M,2020PhRvD.102d3516P}, Gaussianized fields \citep[e.g.,][]{1992MNRAS.254..315W,2009ApJ...698L..90N,2011ApJ...735...32W,2021JCAP...03..070R}, reconstructed density fields \citep[e.g,][]{Eisenstein:2006nk} field-level inference \citep[e.g.,][]{2020JCAP...01..029E,2020JCAP...04..042C,2021JCAP...04..032S}, \resub{Minkowski functionals and other topological descriptors \citep[e.g.,][]{1987ApJ...319....1G,Matsubara:1994wn,1996cceu.conf...45M,Schmalzing:1997cv,SDSS:2003xnk,2014ApJ...796...86P,2022arXiv220308262B,Me94,Me94b,Sousbie:2007pn}}, and beyond), there is little consensus on which have practical utility (with most having been applied only to the dark matter distribution), and few are natural from a theoretical standpoint. An important insight is that the galaxy distribution is simply a set of irregularly arranged point-like particles in three-dimensions; this is mathematically identical to the structure of many terrestrial materials, including atomic
systems, colloids and sphere packing. As such, both scenarios can be treated with the same mathematical formalisms; \textit{i.e.}\ heterogeneous media and statistical mechanical techniques designed to quantify the clustering of particles in materials can be used to provide a practical and well-motivated manner in which to understand the galaxy distribution.

This work considers the application of a number of  
statistical descriptors from the theory of disordered
heterogeneous media to characterize structurally 
the distribution of galaxies, which we treat as  discrete point configuration. We ask two main questions: (1) what can we learn about cosmology through the lens of disordered heterogenous media and statistical mechanics? (2) what condensed matter physics lessons, more broadly,  can we learn from cosmological structures ? As a proof-of-concept, we will consider a number of descriptors \citep[e.g.,][]{peebles80,To02a}, including the two- and three-particle correlation functions, `void' and `particle' nearest-neighbor functions, certain cluster statistics, pair-connectedness functions, percolation properties, and scalar order metrics to quantify the degree of order. We show how their behaviors in the cosmic landscape, probed through \resubtwo{cosmological dark matter simulations}, differs substantially from that expected from a simple Poissonian distribution of points as well as well-known homogeneous models of correlated disordered point patterns, showing that cosmology challenges  general expectations of standard heterogeneous media and statistical mechanical models. In particular, we will find stronger clustering on large-scales, giving an enhancement in the pair correlation and pair-connectedness functions and an excess of large scale voids; these effects arise due to early Universe physics, which source quasi-long-range correlations in the galaxy distribution and create a hyperuniform system. Particular interest will be paid to the question of clustering and phase `percolation'; this is a well-understood phenomenon for many models in condensed matter but can be similarly extended to galaxy distributions, and yields interesting results, The cosmological case will be found to percolate faster, but asymptote to the Poisson case if the density is low, and both scenarios share the same set of critical exponents. Furthermore, we will consider the utility of  descriptors from the theory of disordered heterogeneous media in cosmological settings, quantifying how they can add additional information regarding the early and late Universe, finding that the pair-connectedness function adds significant cosmological information at minimal additional computational cost. \new{We caution that further work will be required before the statistics can be applied to observational data: this must include discussion of redshift-space effects (arising from the conversion of galaxy redshifts to distance, creating anisotropy with respect to the sample line-of-sight), and the dependence of descriptors on galaxy properties, such as luminosities or masses.}

Our conclusions to the above questions asked will be the following: (1) the theory of heterogeneous media and statistical mechanics provides an array of useful tools that can enhance the utility of galaxy survey datasets, strengthening the constraints on cosmological parameters and probing novel features of the distribution, (2) due to its quasi-long-ranged correlations, galaxy samples exhibit very different behavior to most terrestrial media, and thus provides an important sandbox for applying and understanding condensed matter techniques. Whilst we restrict ourselves to galaxy surveys in this work, they are by no means the only cosmological application of such statistics: a number of other phenomena could be described by such approaches. These include two-phase media such as the distribution of cosmic voids (empty regions of gargantuan extent) and the statistics of ionized hydrogen bubbles during the `reionization' phase of the Universe. Such areas provide a bountiful mine from which to derive future work.

\vskip 4pt

The remainder of this work is structured as follows. In \S\ref{sec: statistics}, we provide an overview of the statistics used in this work, before presenting the proposed testing ground (simulated galaxy samples) in \S\ref{sec: galaxies}. Comparison of the statistics on galactic and Poisson data is shown in \S\ref{sec: statistics}, with \S\ref{sec: percolation} providing a discussion of percolation physics in the two systems. Finally, \S\ref{sec: pair-connectedness} considers the utility of a specific statistic, the pair-connectedness function, in cosmological contexts, before we conclude in \S\ref{sec: summary}.

\section{Statistical Descriptors of Point Configurations}\label{sec: statistics}







\begin{figure}
\centering
\begin{minipage}{.46\textwidth}
  \centering
  \includegraphics[width=\textwidth]{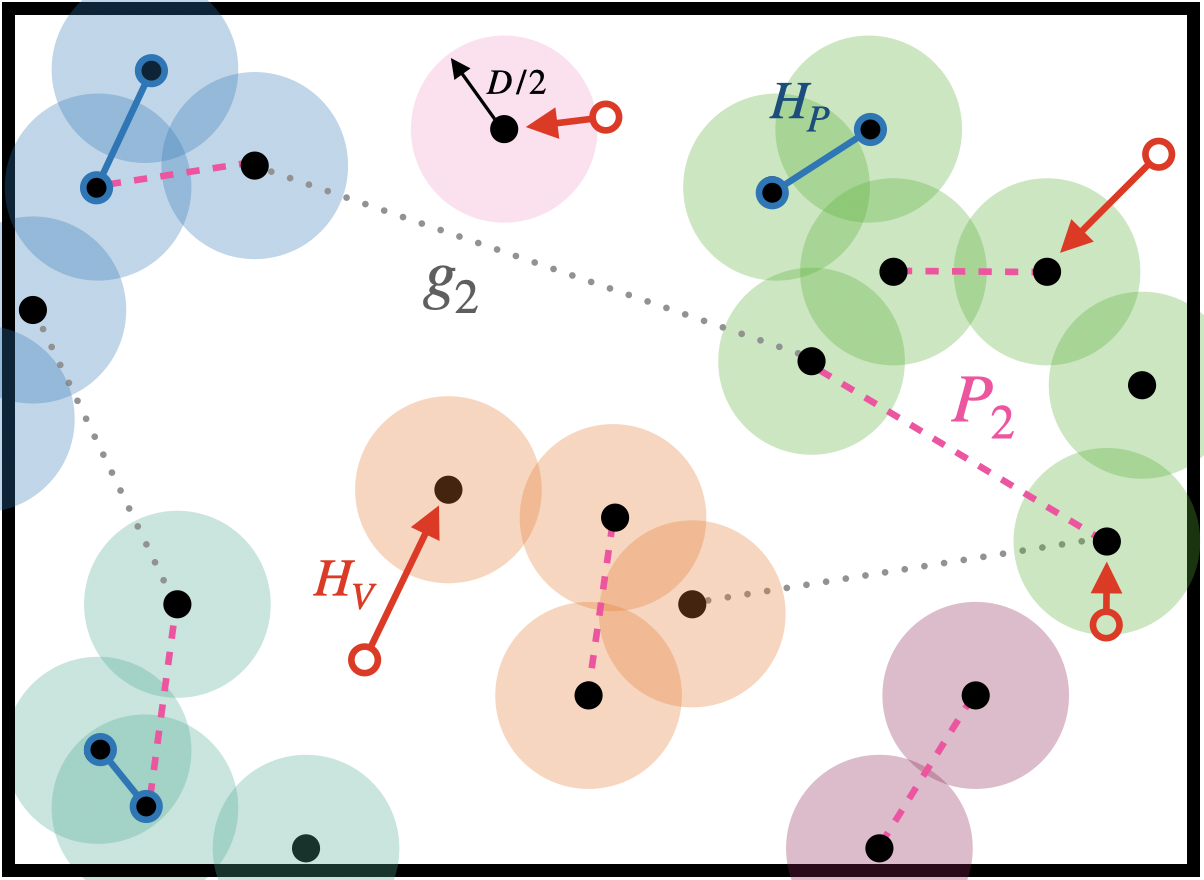}
  \caption{Schematic depicting the various statistical descriptors used in this work. The black points indicate the positions of random particles (here visualized in $\mathbb{R}^2$), with the groups of colored circles (each of diameter $D$) demonstrating the clusters. The pair correlation function, $g_2$, counts pairs of particles belonging both to the same cluster (pink lines), and to different clusters (grey lines), whereas the pair-connectedness function, $P_2$, contains only particles within the same cluster (pink lines). We show also the nearest-neighbor functions: $H_V$ encodes the distance between a randomly positioned point and the nearest galaxy (red arrows), whilst $H_P$ gives the separation between a galaxy and its closest neighbor (blue lines). Further details on these statistics are given in \S\ref{sec: statistics}.
  }
  \label{fig: cartoon}
\end{minipage}%
\hfill
\begin{minipage}{.50\textwidth}
  \centering
    \includegraphics[width=0.95\textwidth]{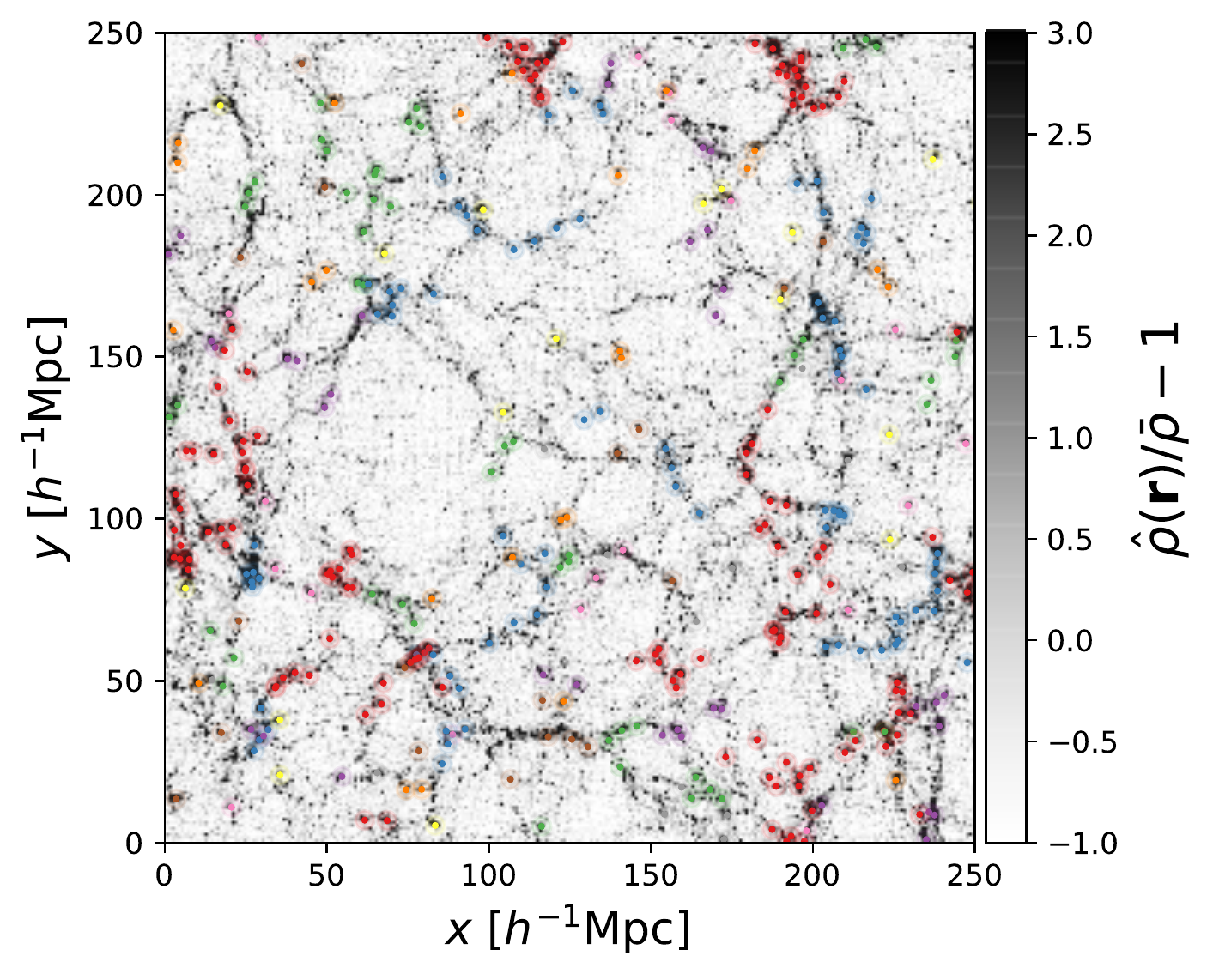}
    \caption{Example of the cosmological simulations used in this work. This shows a $(250\times 250\times 50)h^{-3}\mathrm{Mpc}^3$ slice of the dark matter distribution from a single \textsc{Quijote} simulation, with the colorbar indicating the fractional density of simulation particles in each pixel. The colored points show the positions of \textit{dark matter halos}; these are used as a proxy for galaxies. To facilitate analysis of percolation and connectedness, galaxies are assigned to clusters (indicated by the various colors), via a burning (or `friends-of-friends') algorithm with a separation corresponding to reduced density, $\eta$, here set to $0.2$. Throughout this work, length is given in cosmologists' units of Megaparsecs divided by the reduced expansion rate $h\approx 0.7$), in which the mean pairwise particle separation is $\approx 8\Mpch$.}
    \label{fig: density}
\end{minipage}
\end{figure}

In the section, we define the various statistical descriptors of point configurations that will be used in the remainder of this work. We principally adopt notation from the statistical mechanics community (particularly following \citep{To02a}), though connect this to the cosmological terminology, when relevant. Although we principally work in $\mathbb{R}^3$, most of the following discussion remains relevant in other metric spaces. A schematic illustrating the various statistical descriptors
considered in this work is shown in Fig.\,\ref{fig: cartoon}.

\subsection{Correlation Functions}
The fundamental quantity describing a discrete set of $N$ points in some large region in $\mathbb{R}^d$ of volume $V$ is the $N$-particle probability density functions, $P_N$, which is defined such that $P_N(\vr_1,\cdots,\vr_N)\rmd\vr_1\cdots\vr_N$ is the probability of finding the first particle within $\rmd\vr_1$ of $\vr_1$, the second within $\rmd\vr_2$ of $\vr_2$ \textit{et cetera} \citep[e.g.,][]{peebles80,To02a}. Of more practical use is the $n$-particle probability density function, marginalized over the positions of the other $(N-n)\geq 0$ particles: this is defined as
\beq\label{eq: n-point-pdf}
    \rho_n(\vr_1,\cdots,\vr_n) = \frac{N!}{(N-n)!}\int \rmd\vr_{n+1}\cdots \rmd\vr_{N}\,P_N(\vr_1,\cdots,\vr_n,\vr_{n+1},\cdots,\vr_N),
\eeq
with $\rho_n(\vr_1,\cdots,\vr_n)\rmd\vr_1\cdots\vr_n$ being proportional to the probability of finding one indistinguishable particles within $\rmd\vr_1$ of $\vr_1$ \textit{et cetera}. 

For a statistically homogeneous medium, 
the one-particle density function is a constant, \textit{i.e.} $\rho_1(\vr)\equiv\bar\rho$,  which is the mean {\it number density} (number of points per unit volume in the infinite-volume or `thermodynamic' limit), \resub{commonly labelled $\bar n$ in cosmological contexts (with $\bar\rho$ often used to refer to the mean energy density of the Universe)}. More generally, for statistically homogeneous systems, $\rho_n(\vr_1,\cdots,\vr_n)$
is translationally invariant, enabling us to re-express it as follows:
\begin{eqnarray}
\rho_n(\vr_1,\cdots,\vr_n)={\bar\rho}^n  
g_n(\vr_{12},\cdots, \vr_{1n}),
\label{nbody}
\end{eqnarray}
where $g_n(\vr_{12},\ldots, \vr_{1n})$ is the {\it $n$-particle correlation function} (closely related to the cosmologists' $n$-point correlation function), which depends on the relative positions $\vr_{12},  \cdots$,
where $\vr_{ij} \equiv \vr_j -\vr_i$.
The two-particle or pair correlation function $g_2$ is particularly important in applications, and is schematically illustrated in Fig.\,\ref{fig: cartoon}.
For translationally invariant point configurations without {\it long-range order}, 
$g_n(\vr_{12},\cdots, \vr_{1n}) \rightarrow 1$ when
the points (or `particles') are mutually far from one another, i.e.,  
as $|\vr_{ij}| \rightarrow\infty$ 
($1\leq i < j < n$), $\rho_n(\vr_{1}, \vr_2, \ldots, \vr_{n}) \rightarrow \rho^n$.
Thus, the deviation of $g_n$ from unity  provides a
measure of the degree of spatial correlations (positive or negative) between the particles.
Note that for a translationally invariant {\it Poisson} (spatially uncorrelated) point configurations, 
$g_n=1$ is unity for all values of its argument. 
If the point configuration is in addition statistically isotropic, the functions $g_n$ are invariant under joint rotations of $\vr_{ij}$, such that $g_2$ is a function only of $|\vr_{12}|\equiv r_{12}$, and $g_3$ depends only on $r_{12}$, $r_{13}$ and $r_{23}$.\footnote{\new{Due to the conversion of cosmological redshifts into distances, the observed galaxy density is not isotropic, but distorted along the line-of-sight to the galaxy survey, $\hn$. As such, the pair correlation function depends on an additional angle, \textit{i.e.}\ $g_2(r)\to g_2(r,\hr\cdot\hn)$. We neglect this dependence in this work, but note that it will be important when the statistics discussed herein are applied to observational data.}}


\new{In cosmology it is commonplace to consider not} the probability density function of the full ensemble, but a set of realizations of the microscopic density (often known as the `density field'), each denoted by $\hat\rho(\vr)$. This gives the probability that there is a particle within $\rmd\vr$ of $\vr$ for a specific field (for sufficiently small $\rmd\vr$), and can be represented as a sum of $N$ Dirac deltas: $\hat\rho(\vr) = \sum_{i=1}^N\delta_{\rm D}(\vr-\vx_i)$. Averaging over realizations (denoted by the expectation operator $\mathbb{E}$), we can relate $\hat\rho(\vr)$ to the $n$-particle probability density functions
\beq\label{eq: rho-n-as-av}
    \mathbb{E}\left[\hat\rho(\vr_1)\cdots\hat\rho(\vr_n)\right] \equiv \rho_n(\vr_1,\cdots,\vr_n).
\eeq
If the field is statistically homogeneous, and the volume (\textit{i.e.} $\int \rmd\vr$) sufficiently large, this is equivalent to a spatial average via the ergodic principle. In cosmological contexts, \eqref{eq: rho-n-as-av} is usually adopted, with $\hat\rho(\vr)$ often referred to as $\bar\rho\left[1+\delta(\vr)\right]$ for \textit{overdensity} field $\delta$. Additionally, it is conventional to work with \textit{disconnected} correlation functions, $\xi^{(n)}$, \new{often known as `$n$-point correlation functions': the first few satisfy}
\beq
    \xi^{(2)}(\vr_{12}) &\equiv& \mathbb{E}\left[\delta(\vr_1)\delta(\vr_2)\right] \equiv g_2(\vr_{12}) - 1\\\nonumber
    \xi^{(3)}(\vr_{12},\vr_{13},\vr_{23}) &\equiv& \mathbb{E}\left[\delta(\vr_1)\delta(\vr_2)\delta(\vr_3)\right] \equiv g_3(\vr_{12},\vr_{13},\vr_{23}) - g_2(\vr_{12}) - g_2(\vr_{13}) - g_2(\vr_{23}) + 2
\eeq
\citep[e.g.,][]{peebles80}, and are all zero under Poisson statistics. \new{We will principally work with the full $g_n$ functions in this work, adopting statistical mechanics conventions.}

A particularly  important descriptor is the \textit{structure factor}, $\mathcal{S}(k)$, which is related to the Fourier transform of the total correlation function $h(r)\equiv g_2(r)-1$:
\beq
    \mathcal{S}(\vk) \equiv 1+\bar\rho\,\tilde h(\vk) \equiv 1+\bar\rho\,\int \rmd\vr\,e^{i\vk\cdot\vr}\left[h(\vr)\right].
\eeq
For a Poisson point distribution, $\mathcal{S}(\vk)=1$ for all $\vk$.
The structure factor and the cosmologists' \textit{power spectrum} \cite{peebles80}, $P(\vk)$,
are trivially related to one another via $\mathcal{S}(\vk)={\bar \rho}P(\vk)$. 

The structure factor provides a  useful way to quantify  large-scale (low wavenumber $k\equiv |\vk|$) correlation and fluctuation properties of a point configuration 
and plays a central role in the hyperuniformity concept.
Hyperuniform states of matter are correlated systems that are characterized by an
anomalous suppression of long-wavelength (i.e., large-length-scale) density fluctuations compared
to those found in garden-variety disordered systems, such as ordinary fluids and amorphous solids \cite{To03a,To18a}.
A hyperuniform (or superhomogeneous \cite{Ga02}) many-particle system in $d$-dimensional Euclidean space
$\mathbb{R}^d$ is one in which (normalized)
density fluctuations are completely suppressed at very large length scales,
implying that the structure factor $\mathcal{S}(\vk)$ tends to zero in the infinite-wavelength limit, i.e.,
\begin{equation}
\lim_{|\vk| \to 0}\mathcal{S}(\vk)=0.
\end{equation}
Equivalently, a hyperuniform system is one in which the number variance $\sigma^2_{_N}(R) \equiv \langle N(R)^2 \rangle -\langle N(R) \rangle^2$ of particles within a spherical observation window of radius $R$ grows more slowly than the window volume in the large-$R$ limit, i.e.,
slower than $R^d$.  Typical disordered systems, such
as liquids and structural glasses, have the standard asymptotic volume
scaling $\sigma^2_{N}(R) \sim R^d$ and hence are not hyperuniform. For general translationally invariant
point configation in $\mathbb{R}^d$, the local number variance $\sigma^2_N(R)$ is determined exactly by the pair statistics \cite{To03a}:
\begin{eqnarray}\label{eq: sigmaN-def}
\sigma_N^2(R)&=& \bar{\rho} v_1(R)\Big[1+\rho  \int_{\mathbb{R}^d} h(\vr) 
\alpha(r;R) \rmd\vr\Big] \nonumber \\
&=&
 \bar\rho v_1(R)\Big[\frac{1}{(2\pi)^d} \int_{\mathbb{R}^d} \mathcal{S}(\vk)
{\tilde \alpha}(k;R) \rmd\vk\Big],
\label{local}
\end{eqnarray}
where
$v_1(R) =\pi^{d/2} R^d/\Gamma(1+d/2$
is the volume of a $d$-dimensional sphere of radius $R$, and
$\alpha(r;R)$ is the {\it scaled intersection volume}, the ratio of the intersection volume of two spherical windows
of radius $R$ whose centers are separated by a distance $r$ to the volume of
a spherical window, known analytically in any space dimension \cite{To02a,To06b}. Its Fourier transform
is the nonnegative function given by
\begin{equation}
{\tilde \alpha}(k;R)= 2^d \pi^{d/2} \Gamma(1+d/2)\frac{[J_{d/2}(kR)]^2}{k^d},
\label{alpha-k}
\end{equation}
where $J_{\nu}(x)$ is the Bessel function of order $\nu$.

Consider translationally invariant point configurations that are characterized by a structure factor with a radial power-law form in the
vicinity of the origin, i.e.,
\begin{align}
  \mathcal{S}(\vk) \sim |\vk|^\alpha \quad \text{for } |\vk|\to 0.
  \label{eq:Sk-scaling}
\end{align}
For hyperuniform systems, the exponent $\alpha$ is positive ($\alpha>0$)
and its value  determines three hyperuniformity classes corresponding to different large-$R$ scaling behaviors  of the
number variance \cite{To03a,Za09,To18a}:
\begin{align}
  \sigma^2(R)&\sim\left\{\begin{array}{l l}
    R^{d-1},      &\alpha>1 \text{ (class I)}\\
    R^{d-1}\ln R, &\alpha=1 \text{ (class II)\,.}\\
    R^{d-\alpha}, &\alpha<1 \text{ (class III)}
  \end{array}\right. 
  \label{eq:sigma-scaling}
\end{align}
Classes I and III describe the
strongest and weakest forms of hyperuniformity, respectively.
States of matter that belong to class I include all perfect
crystals \cite{To03a,Za09}, many perfect quasicrystals \cite{Za09,Li17a,Og17}, and
`randomly' perturbed crystal structures  \cite{Ga04b,Ga04,Ga08,Ki18a},
classical disordered ground states of matter \cite{To03a,Uc04b,To15} as well as
systems out of equilibrium \cite{Zh16a,Le19a}. Class~II hyperuniform systems include some
quasicrystals~\cite{Og17}, the positions of the prime numbers \cite{To19},
and many disordered classical ~\cite{Do05d,Za11a,Ji11c,At16a,Zh16a}
and quantum ~\cite{Fe56,Re67,To08b} states of matter. Examples of class~III hyperuniform systems include
classical disordered ground states~\cite{Za11b}, random organization models~\cite{He15} and
perfect glasses~\cite{Zh16a}. 
 Certain disordered hyperuniform systems are poised at an `inverted' critical point in which the volume integral of the total correlation function $h(\vr)$ is quasi-long-ranged but its volume integral is bounded \cite{To03a,To18a}.

By contrast, for any nonhyperuniform system, the local variance has the following large-$R$ scaling behaviors \cite{To21a}:
\begin{align}  
\sigma^2(R) \sim 
\begin{cases}
R^{d}, & \alpha =0 \quad \text{(typical nonhyperuniform)}\\
R^{d-\alpha}, & -d< \alpha < 0 \quad \text{(anti-hyperuniform)}.\\
\end{cases}
\label{sigma-nonhyper}
\end{align}
For a  `typical' nonhyperuniform system, the structure factor $\mathcal{S}(0)$ is bounded  \cite{To18a}. In {\it anti-hyperuniform} systems,
$\mathcal{S}(0)$ is unbounded, i.e.,
\begin{equation}
\lim_{|\vk| \to 0} \mathcal{S}(\vk)=+\infty,
\label{antihyper}
\end{equation}
and hence  are diametrically opposite to hyperuniform systems.
Anti-hyperuniform systems include fractals, systems at thermal critical points (e.g., liquid-vapor and magnetic critical points) \cite{Wi65,Ka66,Fi67,Wi74,Bi92}
as well as certain substitution tilings \cite{Og19}.

\subsection{Order Metric}\label{subsec: disorder}

\resub{
Given the richness of the spectrum of possible  microstructures that can arise in condensed phase systems,
an   outstanding challenging task has been  the quantification
of their degree  order/disorder. Scalar order/disorder metrics
have been profitably employed to quantify the degree of order in many-particle systems, including sphere packings; see \cite{To02a}\,\&\,\cite{To18b}
and references therein. Any scalar order metric $\Psi({\bf R})$  is a
well-defined nonnegative scalar function of a
many-particle configuration ${\bf R}$ and  if, for any two configurations ${\bf R}_A$ 
and ${\bf R}_B$, $\Psi({\bf R}_A)>\Psi({\bf R}_B)$, we say that configuration ${\bf R}_A$ is
to be considered more ordered than configuration ${\bf R}_B$.
It has been suggested that a good scalar order metric should have
the following additional properties \cite{Ka02d}: (1)
sensitivity to any type of ordering without bias toward any reference system; 
(2) ability to reflect the hierarchy of ordering between prototypical systems given by
common physical intuition (e.g., perfect crystals with
high symmetry should be highly ordered, followed by
quasicrystals, correlated disordered packings without
long-range order, and finally spatially uncorrelated or
Poisson distributed particles); (3) capacity to detect order at any length scale; 
and (4) incorporation of both the
variety of local coordination patterns and the spatial distribution of such patterns.}

\resub{The recently introduced $\tau$ order metric \cite{To15} fulfills these requirements
and has been fruitfully employed to characterize the degree of order 
across length scales of a diverse set of \resubtwo{disordered media \cite{To15,At16b,Kl19a,martelli+17}}.
This metric, which we compute here  for the first time for the galaxies, is defined as}
\begin{eqnarray}
    \tau &= &\frac{1}{D^d}\int_{\mathbb{R}^d}\rmd\vr\,h^2(\vr) \nonumber\\
 &=&   \frac{1}{(2\pi)^d D^d{\bar \rho}^2}\int_{\mathbb{R}^d} \rmd\vk \,[\mathcal{S}({\vk})-1]^2,
\label{tau}    
\end{eqnarray}
where $D$ is a characteristic `microscopic' length scale. This scalar metric measures deviations of two-particle statistics from that of the Poisson distribution. Since both positive and negative correlations contribute to the integral,
due to the fact that $h(\vr)$ is squared, $\tau$
measures the degree of translational order
across length scales. It clearly vanishes for the uncorrelated Poisson distribution, diverges for an infinite crystal and is a positive bounded number for correlated
disordered systems without long-range order
(i.e., Bragg peaks). \resub{It is interesting to note
that the $\tau$ order metric is closely related
to the negative of the excess two-particle entropy of the system \cite{Lo17}.}

\subsection{Nearest-Neighbor Functions}\label{subsec: nearest-neighbor}
Another well-known set of statistical descriptors that
arise in rigorous bounds on the macroscopic physical
properties of disordered heterogeneous media, such as
suspensions of spheres, and employed 
in the statistical mechanics of many-particle systems  are nearest-neighbor functions \cite{Torquato90,To02a,To08b}. There two types of such functions: `void' and
`particle' quantities. The void and particle nearest-neighbor probability
density functions $H_{V}(r)$ and $H_{P}(r)$, respectively,
are defined as follows:
\begin{eqnarray}
\begin{array}{ccp{3.7in}}
H_{V}(r)\,\rmd r & = & Probability that a point of the point configuration
lies at a distance between $r$ and $r + \rmd r$ from an arbitrary
point in the space. 
\end{array} 
\label{def-Hv}
\end{eqnarray}
\begin{eqnarray}
\begin{array}{ccp{3.7in}}
H_{P}(r)\,\rmd r & = & Probability that a point of the point configuration
lies at a distance between $r$ and $r + \rmd r$ from another 
point of the point configuration.
\end{array} 
\label{def-Hp}
\end{eqnarray}
The associated dimensionless `exclusion' probabilities $E_{V}(r)$ and  $E_{P}(r)$ are defined as follows:
\begin{eqnarray}
\begin{array}{ccp{3.7in}}
E_{V}(r) & = & Probability of finding a spherical cavity
of radius $r$ empty of any points in the point configuration.
\end{array}
\label{def-Ev}
\end{eqnarray}
\begin{eqnarray}
\begin{array}{ccp{3.7in}}
E_{P}(r) & = & Probability of finding a spherical cavity
of radius $r$ centered at an arbitrary point of the point configuration
empty of any other points.
\end{array}
\label{def-Ep}
\end{eqnarray}
It follows that the exclusion probabilities are {\it complementary cumulative distribution functions} associated with the density functions and thus are related to the latter via
\begin{eqnarray}
E_{V}(r) = 1 - \int_{0}^{r} H_{V}(x) \, \rmd x, \qquad H_V(r) = -\partial_r E_V(r)
\label{Ev-cum}
\end{eqnarray}
and
\begin{eqnarray}
E_{P}(r) = 1 - \int_{0}^{r} H_{P}(x) \, \rmd x, \qquad H_P(r) = -\partial_r E_P(r).
\label{Ep-cum}
\end{eqnarray}
The moments of $H_V(r)$ and $H_P(r)$, defined by
\begin{equation}
\ell_V^{(k)}=\int_0^\infty r^k H_V(r) dr, 
\label{mom-V}
\end{equation}
\begin{equation}
\ell_P^{(k)}=\int_0^\infty r^k H_P(r) dr,
\label{mom-P}
\end{equation}
are particularly useful integral nearest-neighbor measures,
with the $k=1$ version of the latter representing the {\it mean nearest-neighbor distance between particles.} \new{The void nearest neighbor function, $H_V$, has received some attention in cosmology, \resub{both historically \citep{1984ApJ...287L..59R,White:1979kp,1989A&A...220....1B}}, and in recent works, in particular via the `$k$NN' statistics \citep{2021MNRAS.500.5479B,2021MNRAS.504.2911B,2022MNRAS.511.2765B,2022MNRAS.514.3828W}, generalizing the above to the $k$-th nearest neighbor. This has been shown to yield strong constraints on cosmological parameters (cf.\,\S\ref{sec: pair-connectedness}), and can be modeled semi-analytically.} It is noteworthy that
$k$NN statistics and related quantities have been studied and fruitfully applied in the field of  statistical mechanics \cite{Ve75,Zi77,Tr98b,To21b}.

Both the void and particle nearest-neighbor functions generally involve integrals over all the $n$-particle correlation functions, $\{g_n\}$ ($n=2,3,4,\ldots)$ \new{\citep{Torquato90,2021MNRAS.500.5479B}}. 
While the void and particle nearest neighbor functions
are identical to one another for a Poisson point configuration (e.g., $E_V(r)=E_P(r)=\exp(-4\pi r^3{\bar \rho}/3$) in three dimensions), they are generally
different from one another for correlated systems,
as manifested by their different series representations \cite{Torquato90}. Both the void and particle quantities
arise in rigorous bounds on the effective transport
and mechanical properties of heterogeneous media \cite{To02a}. It is noteworthy that the void nearest-neighbor functions play a deep role in the
covering problem of discrete geometry \cite{To10d}.
The covering problem asks for the point
configuration that minimizes the radius of overlapping spheres circumscribed around each of the points required to cover $d$-dimensional Euclidean space $\mathbb{R}^d$ \cite{Co93}. The above void statistic \new{also} bears some similarities to the `void size function' used in cosmology \citep[e.g.,][]{Sheth:2003py,2019BAAS...51c..40P}. The latter quantity is usually constructed from smoothed density fields, with voids identified as spherical regions with a mean density below some critical threshold, usually $30\%$ of the system's mean density. This differs from $H_V$ in two key ways: (a) the voids defined by \eqref{def-Hv} contain no particles, thus are equivalent to requiring a critical density of zero, (b) the cosmologists’ void contains no sub-voids: \textit{i.e.}\ any empty area of space within a void cannot be classified as a smaller void, unlike for $H_V$.


\subsection{Clustering and Connectedness Functions}

To quantify the geometrical and topological properties of the class of disordered  heterogeneous media consisting of particles distributed throughout a matrix phase, it is often useful to statistically characterize \textit{particle clusters}  that
are defined according to some connectivity criterion \citep[e.g.,][]{Coniglio_1977,St84,torquato1988two,To02a,jiao+09,torquato12} For point (zero-dimensional) particles, this can be achieved by circumscribing
each point by spheres of diameter $D$, which
generally may overlap with one another. Such a decoration
of the points by possibly overlapping spheres divides the space into two disjoint regions or `phases,' encompassing points that do and do not lie within a distance $D/2$ of at least one point. Two spheres are deemed
to be connected if they overlap. Defining the {\it reduced density}  $\eta \equiv \bar \rho \pi D^3/6$ (in $\mathbb{R}^3$), it is clear that as 
the diameter $D$ at fixed mean density $\bar \rho$ increases from zero, $\eta$ and the
fraction of space occupied by the spheres will
increase and clusters of various sizes will form
and grow \citep{To02a}. 

Once clusters have been identifed, one can determine the \textit{pair-connectedness function}, $P_2(r,\eta)$, where $P_2(r,\eta)\times 4\pi r^2dr$ is the conditional probability of finding a particle in a shell of radius $dr$ at radial distance $r$ from another particle in the same cluster (assuming statistical homogeneity and isotropy). Equivalently, this quantity gives the probability that there exists a path from the first to the second point that never leaves the particle phase, \textit{i.e.} one that is always within a distance $D/2$ of at least one particle (cf.\,Fig.\,\ref{fig: cartoon}).\footnote{Formally, this can be defined as $P_2(r,\eta)\equiv\mathbb{E}\left[\hat\rho(\vx)\hat\rho(\vx+\vr)\hat\Phi(\vx,\vx+\vr,\eta)\right]$, for clustering function 
\beq\label{eq: P2-def}
    \Phi(\vx,\vx+\vr,\eta) = \begin{cases} 1 & \text{if}\,\,\exists\,\,\vec\Gamma:[0,1]\to\mathbb{R}^3\,\text{ s.t. }\,\left\{\vec\Gamma(0)=\vx,\quad \vec\Gamma(1)=\vx+\vr,\quad \int_0^1\rmd\gamma\,\varphi_D(\vec\Gamma(\gamma))\geq \ell(\vec\Gamma)\right\} \\ 
    0 & \text{else},\end{cases}
\eeq
where we consider all paths $\vec \Gamma$ connecting $\vx$ and $\vx+\vr$ for which the integral of $\varphi(\vr)$ (defined as the particle phase, \textit{i.e.} $\varphi(\vx) = \Theta_H\left[\int \rmd\vy\,\hat\rho(\vy)\Theta_H(D-|\vx-\vy|)\right]$ for Heaviside $\Theta_H$) is at least the line length $\ell(\vec\Gamma)$, \textit{i.e.} those passing only through connected regions.}  This is the \textit{connected} contribution to the full pair correlation function, $g_2(r)$:
\beq
    g_2(r) \equiv P_2(r,\eta) + B_2(r,\eta),
\eeq
where the \textit{pair-blocking function} $B_2(r,\eta)$ gives the correlation between pairs of particles which do not lie in the same cluster. For $r<D(\eta)$, any pair of points must be within the same cluster, thus $P_2(r<D,\eta) = g_2(r)$.

A related quantity is the \textit{direct-connectedness function} $C_2(r,\eta)$ (also known as the non-nodal correlation function) \citep{Coniglio_1977}. This is the probability that two points separated by a distance $r$ are connected by a path through the set of random particles that does not involve nodes (\textit{i.e.} one that cannot be broken by a single cut, as in Fig.\,\ref{fig: cartoon}). A general path between two points contains either zero or at least one node: this permits the \textit{Ornstein-Zernike} (OZ) decomposition \citep{Coniglio_1977,St84}
\beq\label{eq: OZ-Poisson}
    P_2(\vr_{12},\eta) = C_2(\vr_{12},\eta) + \bar \rho\int \rmd\vr_3\,C_2(\vr_{13},\eta)P_2(\vr_{32},\eta),
\eeq
labelling $\vr_{ij}=\vr_i-\vr_j$. The first quantity on the RHS contains paths with no nodes, \resub{thus involves the first factor of $C_2$}, whilst for the second, we integrate over the position of the node closest to $\vr_1$ (assuming a statistically homogeneous field $\bar\rho$), noting that the path from $\vr_1$ to $\vr_3$ contains no nodes by definition, \resub{yielding another function of $C_2$. Finally, the path from $\vr_3$ to $\vr_2$ can contain nodes, leading to the final factor of $P_2$}. In Fourier-space this gives a simple relation between the pair-connectedness and direct-connectedness functions:
\beq\label{eq: P2-C2-relation}
   \tilde C_2(k,\eta) = \frac{\tilde P_2(k,\eta)}{1+\bar\rho\,\tilde P_2(k,\eta)}, \qquad \tilde P_2(k,\eta) = \frac{\tilde C_2(k,\eta)}{1-\bar\rho\,\tilde C_2(k,\eta)}.
\eeq

\resub{To gain intuition for the pair-connectedness function (and related statistics), it is instructive to consider its form for a Poissonian system \resub{(noting that there is no Gaussian limit, given that we are dealing with discrete systems)}. At low densities, it can be computed as a perturbation series in $\bar\rho$ (or, more strictly, in $\eta/\eta_c$), first considering pairs of particles that are directly linked by the covered phase (\textit{i.e.}\ their centers lie within $D(\eta)$), then moving to pairs linked via a third particle and so on. This leads to the decomposition
\beq\label{eq: P2-poiss}
    P_2^{\rm Poiss}(\vr_{12},\eta) &=& \Theta_D(\vr_{12})+\bar \rho\,\left[1-\Theta_D(\vr_{12})\right]\int \rmd\vr_3\,\Theta_D(\vr_{13})\Theta_D(\vr_{32})\\\nonumber
    &&\,+\,\bar \rho^2\left[1-\Theta_D(\vr_{12})\right]\int \rmd\vr_3\rm\rmd\vr_4\,\Theta_D(\vr_{13})\Theta_D(\vr_{34})\Theta_D(\vr_{24})\left[1-\Theta_D(\vr_{14})\right]\left[1-\Theta_D(\vr_{23})\right]+\ldots,
\eeq
where the Heaviside function $\Theta_D(\vr)\equiv \Theta_H(r-D)$ selects pairs with separations below $D$. In this expansion, successive terms integrate over progressively more particle positions with, for example, the second term averaging over the position of $\vr_3$, which must be within a distance of $D$ from both $\vr_1$ and $\vr_2$. As such, this expression is difficult to compute beyond second order (which is convolutional) and thus rarely used in practice, unless $\eta\ll\eta_c$ and we restrict to small scales. In practice, approximate treatments are usually adopted, such as via the OZ equation \eqref{eq: OZ-Poisson} combined with heuristic `closure' relations such as the Percus-Yevick form \citep{percus-yevick}. These give accurate predictions for $P_2(r,\eta)$ in low density regimes at relatively small $r$. This stands in contrast to the case familiar from cosmology, when the modeling of $g_2(r)$ becomes progressively more accurate as $r$ increases.}

\resub{For a general system, a similar decomposition to \eqref{eq: P2-poiss} is possible, and takes the form:
\beq
    P_2(\vr_{12},\eta) &=& g_2(\vr_{12})\Theta_D(\vr_{12})+\bar \rho\,\left[1-\Theta_D(\vr_{12})\right]\int \rmd\vr_3\,g_3(\vr_{13},\vr_{32})\Theta_D(\vr_{13})\Theta_D(\vr_{32})\\\nonumber
    &&\,+\bar \rho^2\,\left[1-\Theta_D(\vr_{12})\right]\int \rmd\vr_3\rmd\vr_4\,g_4(\vr_{13},\vr_{34},\vr_{42})\Theta_D(\vr_{13})\Theta_D(\vr_{34})\Theta_D(\vr_{24})\left[\Theta_D(\vr_{14})\right]\left[1-\Theta_D(\vr_{23})\right]+\ldots.
\eeq
In this case, the expansion depends on the correlation functions $g_n$, since there exists background correlations in addition to that induced by the circumscribed spheres around points. As expected, this implies that $P_2(r<D,\eta)=g_2(r)$, with the $n$-th order term involving correlators of the form $g_{n+2}$. Formally, this expression may be extended to all orders, via the relation
\beq\label{eq: P2-formal-expan}
	\bar\rho^2P_2(r,\eta) = \prod_{k=2}^{\infty}\bar\rho^k \int \left(\prod_{i=1}^k \rmd\vr_i\right)\delta_{\rm D}(|\vr_1-\vr_k|-r)g_k(\vr_1,\ldots,\vr_k)\prod_{i=1}^{k-1}\Theta_D(\vr_{i(i+1)})\prod_{m=1}^{k-2}\left[\prod_{j=m+2}^{k}(1-\Theta_D(\vr_{mj}))\right].
\eeq
Whilst this form is not particularly useful for analytic treatments, due to the difficulty inherent in performing the high-dimensional integrals present for $k>3$ it illustrates how the pair-connectedness function is comprised of all possible correlation functions, and thereby partly resums information found at all orders. \resub{At very low densities, $P_2(r)$ tends to $\Theta_D(r)g_2(r)$,} thus we do not expect this statistic to add information; however, as $\eta$ increases, the fraction of information contributed by the higher-order $g_n$ increases, until the system becomes non-perturbative at $\eta\approx\eta_c$ \resub{(whence the notion of connectedness breaks down)}. One may ask whether an OZ-like equation with some closure relation can be used to provide an approximate analytic form for $P_2(r,\eta)$ in the general case. Unfortunately, this is far from trivial, since any scheme only involving $g_2$ will miss any contributions to $P_2$ from $g_3$ and above, which are of particular cosmological interest, especially when performing parameter inference in conjunction with $g_2$ (as is the case below). We leave further treatment of this problem to future work.}

\subsection{Continuum Percolation}\label{subsec: percolation-stats}
Percolation describes the appearance of a \textit{phase transition} in the system, which, in the above case, corresponds to the emergence of long-range connectivity in the point cloud due to arbitrarily large clusters of points, again defined by spheres of some diameter. This phenomenon is of relevance in a wide variety of physical settings such as the transport of fluid in porous systems, the appearance of fractures in geological formations, spread of diseases, and the collapse of gas into stars \citep{Ha57,To02a,torquato1988two,Sa03,torquato12}. This crossover from non-percolating clusters  to the appearance of the incipient  sample-spanning cluster (infinite in the thermodynamic limit) is characterized by a \textit{critical reduced density} (also known as a percolation density), $\eta_c={\bar \rho} \pi D_c/6$ (in $\mathbb{R}^3$), where $D_c$ is the critical sphere diameter at fixed  mean density $\bar \rho$ with the system said to have percolated for $\eta>\eta_c$. For a Poissonian system in $\mathbb{R}^3$, numerical simulations find $\eta_c\approx 0.34$ \citep{To02a}, implying that the connected phase fills about $29\%$ of the space.
In addition to the connectedness functions described above, percolation theory utilizes a number of other statistical descriptors, which we outline below.

The \textit{mean cluster size}, $S(\eta)$, gives a simple manner in which to characterize clustering and  percolation at some reduced density $\eta$ \citep[e.g.,][]{Coniglio_1977}. This quantity is simply the mean number of in a cluster containing a randomly chosen particle. It is directly
related to the pair-connectedness function by the following relation:
\beq
    S(\eta) = 1+\bar\rho\int \rmd\vr\,P_2(\vr,\eta) = \left[1-\bar\rho\tilde C_2(0,\eta)\right]^{-1},
\eeq
where the second equality follows from a Fourier-transform and using \eqref{eq: P2-C2-relation}. When $\eta>\eta_c$, clusters of infinite extent appear, thus $S(\eta)\to\infty$, and the volume integral of the pair-connectedness function diverges. Given a form for $C_2(r,\eta)$, the second relation provides a useful manner in which to estimate $\eta_c$, by solving $\bar\rho\tilde C_2(0,\eta_c) = 1$. Furthermore, the behavior close to the phase transition can be expressed in terms of \textit{critical exponents} of the field: in particular, $S(\eta)\sim (\eta-\eta_c)^{-\gamma}$ and $C_2(r,\eta)\sim r^{-\alpha}$ at large $r$, for $\eta\to\eta_c^{-}$, where $\alpha$ and $\gamma$ are found to be universal for a broad class of physical models \citep{lee-torquato90,torquato12}.

The mean cluster size can also be written in terms of so-called $s$-mer cluster statistics for densities below the percolation threshold $\eta_c$ \citep{lee-torquato88,To02a} , namely,
\beq\label{eq: S-eta-def}
    S(\eta) = \frac{\displaystyle \sum_{s=1}^\infty s^2 n_s}{\displaystyle \sum_{s=1}^\infty s n_s}, \qquad \eta <\eta_c.
    \label{S-ns}
\eeq
Here $n_s$ is the average number of $s$-mers, clusters containing $s$ particles, per unit number of particles. This representation will be employed to estimate $S(\eta)$ from simulations. 

For a finite (aperiodic) system of size $L$, percolation may be studied by considering the existence of \textit{sample spanning clusters}, \textit{i.e.} clusters of connected points which reach from the top to the bottom of the system (in some dimension). In the $L\to\infty$ limit, these clusters will appear only for $\eta>\eta_c$: for finite systems, the behavior can be characterized using the \textit{percolation probability}, $\Pi(\eta,L)$, which is the probability that a realization of size $L$ will contain a sample-spanning cluster. This will be used to compute the percolation threshold, $\eta_c$, in \S\ref{sec: percolation}, via a finite-scaling analysis. If such a cluster exists, its mass (\textit{i.e.} the number of constituent particles, denoted $M(L,\eta)$) can be used to ascertain the effective mass \textit{fractal dimension}, $d_F$, of the field. In particular:
\beq\label{eq: M-L-scaling}
    M(L,\eta)\sim L^{d_F},\qquad \eta\to\eta_c^{-}
\eeq
where $d_F\approx 2.42$ for Poisson systems, and any other processes in the same universality class \citep[cf.,][]{To02a,gabrielli05,baryshev05}.

\section{Galaxy Surveys as Point Clouds}\label{sec: galaxies}



Although the statistics described in \S\ref{sec: statistics} have been principally applied to study the properties of physical materials, their applicability extends far beyond the terrestrial regime. In this work, we consider their application to spectroscopic galaxy surveys, such as those of the upcoming Dark Energy Spectroscopic Instrument (DESI) and Euclid projects \citep{desi16,euclid11}. Such projects will measure the angular positions and redshifts of $\mathcal{O}(10^7)$ bright galaxies from ground- and space-based telescopes, providing a three-dimensional map of the Universe with unprecedented resolution. Fundamentally, galaxy surveys measure a set of $N$ galaxy positions with some associated weights, representing experimental effects. In many typical analyses \citep[e.g.,][]{tegmark98,boss17,philcox22}, these are assigned to some coarse grid in $\mathbb{R}^3$, and the associated field taken to be an inhomogeneous Poisson sample of an underlying continuous field. This is itself modelled as a non-linear transformation of the underlying \textit{dark matter} distribution, whose correlation functions (particularly $g_2$ and $g_3$) encode early Universe physics, with the field obeying Gaussian statistics on sufficiently large scales, before a perturbative (and well understood, \citep[e.g.,][]{baumann12}) regime takes hold. Explicitly, the microscopic density $\hat\rho(\vx)$ satisfies:
\beq\label{eq: poiss-gauss}
    \hat\rho(\vx) \sim \mathrm{Poisson}\left(\bar\rho\left[1+\delta_{g}(\vx)\right]\right), \qquad \tilde{\delta}_g(\vk) \sim \mathcal{N}(0,\tilde h_{2,g}(k))
\eeq
on sufficiently large scales, where $\delta_g$ is a normally-distributed continuous background field with variance $\tilde h_{2,g}(k)$. On small, non-linear, scales, a variety of galaxy formation processes become important and the above approach is known to be insufficient. This has led to a flurry of interest in additional statistics beyond the simple correlation functions.

An alternative to the standard approach is to consider the point cloud traced by the galaxies as the fundamental object, facilitating direct application of the clustering techniques described in \S\ref{sec: statistics}. Rather than working with observational data directly, this will work will make use of simulated data, drawn from the publicly available \textsc{Quijote} suite \citep{quijote}, which is a collection of $40\,000$ realizations of the Universe, each contained within a cubic volume of size $L=1000\Mpch$.\footnote{Following cosmologists convention, we work in $\Mpch$ units, where $1\,\mathrm{Mpc}\equiv10^6\,\mathrm{parsec}$ and $h^{-1}\approx 1.4$ is used to remove a leading scaling.} In particular, we use dark matter simulations that have been evolved down to redshift zero (today), and contain a set of $\sim 10^5$ \textit{dark matter halos}: spheroidal agglomerations of matter in which galaxies are known to form.\footnote{In this work, we use only halos containing at least 64 dark matter particles to avoid discreteness effects; these have masses \resubtwo{$M\gtrsim 5 \times 10^{12}h^{-1}M_\odot$ in our baseline simulations}.}

Rather than dealing with the complexities of assigning galaxies to dark matter halos as a function of their mass (for example using a halo occupation distribution \citep{zheng05}), we use the positions of the dark matter halos as a direct proxy for the galaxy positions, which is sufficient for this initial study. \resub{As such, we do not require the Poisson-Gaussian assumptions of \eqref{eq: poiss-gauss}, and will utilize the galaxy catalog only as a discrete point cloud.} In most scenarios, we will use the `galaxy' catalogs extracted from $1000$ \textsc{Quijote} \resubtwo{`High--Resolution'} simulations, each run with the same underlying physical model, but with varying realizations of the (stochastic) initial conditions. A section of a typical simulation is shown in Fig.\,\ref{fig: density}. \resubtwo{We caution that these simulations do not fully represent observational data, in particular due to their limited mass resolution and lack of (magneto-)hydrodynamic effects. However, their simplified nature makes them ideal for the proof-of-concept study considered herein, since it allows for a large number of simulations (and thus determination of accurate covariances). Further work will necessarily require application of the above tools to higher-resolution simulations, though these are fewer in number.}


\section{Phenomenological Clustering Statistics}







\begin{figure}
    \centering
    \subfloat[Pair Correlation Function]{%
      \includegraphics[width=0.48\textwidth]{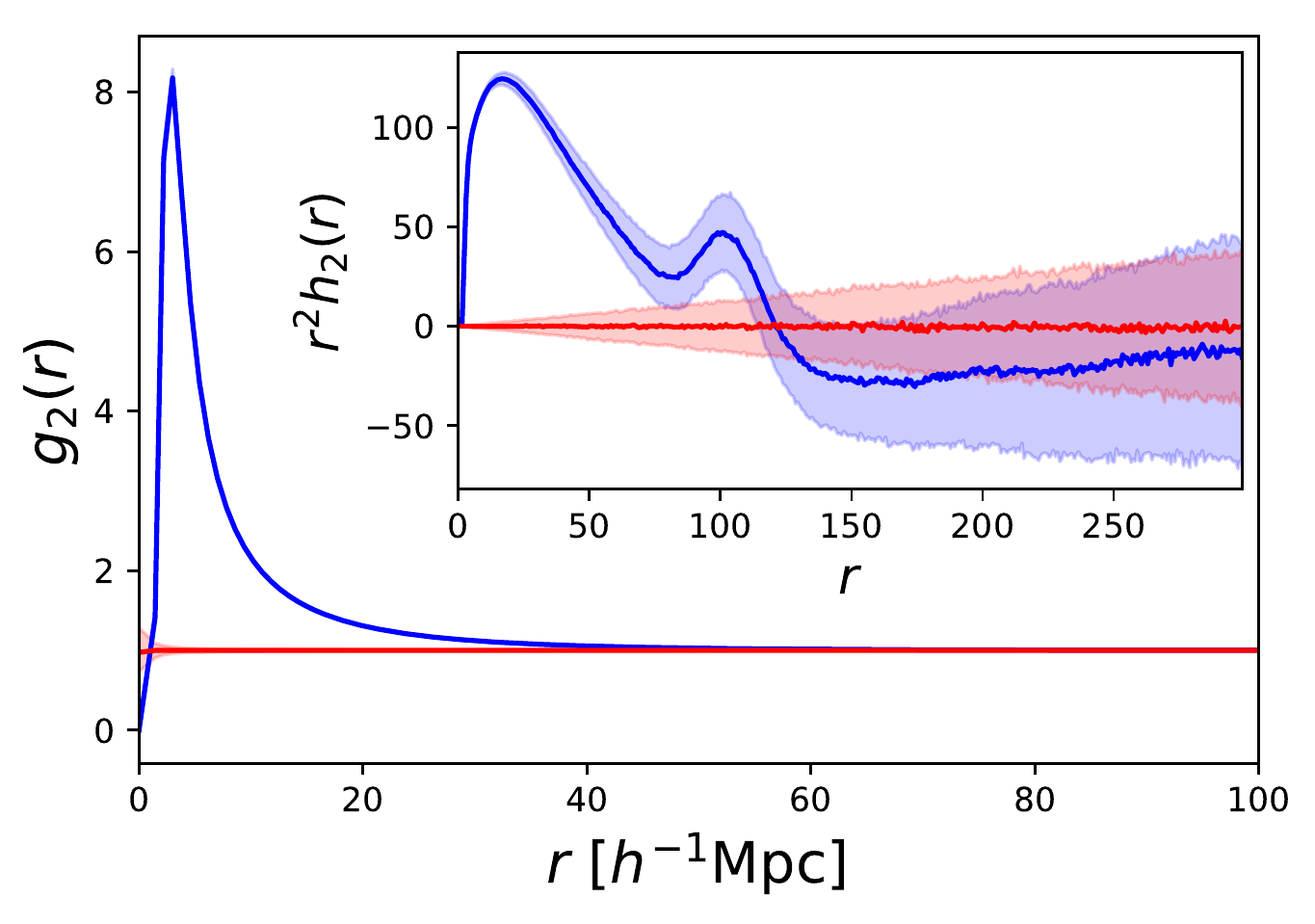}\label{subfig: g2}
    }
    \hfill
    \subfloat[Structure Factor]{%
      \includegraphics[width=0.48\textwidth]{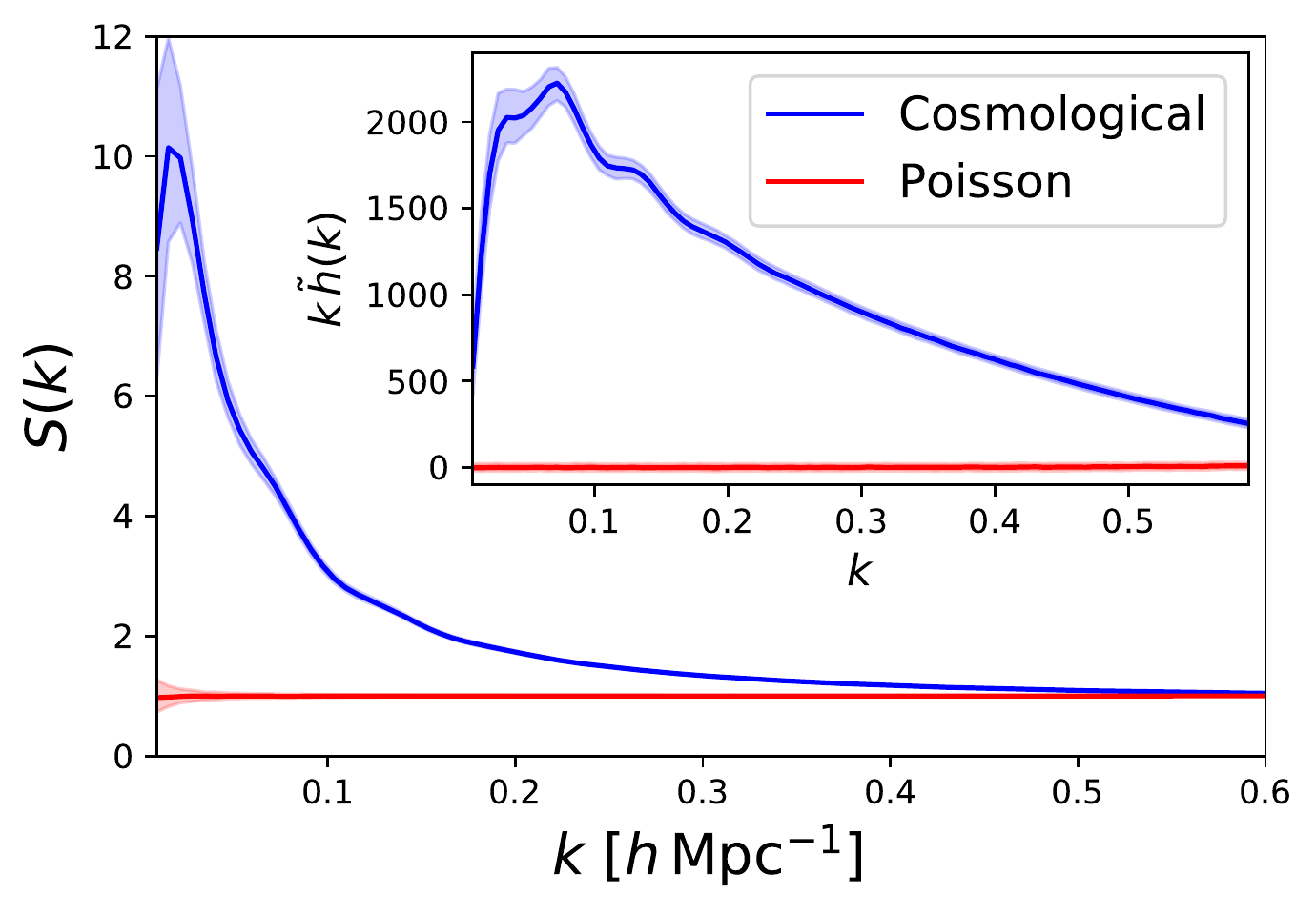}\label{subfig: Sk}
    }
    \caption{Measurements of the pair correlation function and structure factor from 1000 \textsc{Quijote} cosmological simulations (blue) alongside 1000 Poisson realizations (red). Both sets of simulations have the same number density ($\approx 1\times 10^{-4}h^3\mathrm{Mpc}^{-3}$) and a volume of $1h^{-3}\mathrm{Gpc}^3$, with a mean \resub{pairwise particle} separation of $\approx 8\Mpch$. The shaded regions show the statistical variance between realizations (which grows large on large scales), and we note that the two statistics are related by a Fourier transform. In Fig.\,\ref{subfig: g2}, the inset shows the pair correlation function in the typical cosmologists' normalization, plotting $r^2h_2(r)\equiv r^2[g_2(r)-1]$; this clearly brings out the structure imprinted by early Universe physics. At large $r$, we find $h_2(r)\sim r^{-(3+n_s)}$, where $n_s\approx 0.96$: this indicates the presence of quasi-long-range correlations. Similarly, the inset of Fig.\,\ref{subfig: Sk} shows $k\tilde h(k)$, equal to the cosmologists' $kP(k)$ (with a slope of $k^{n_s}$ on large scales). The oscillatory features  at $k\sim 0.1\hMpc$ arise from acoustic waves in the early Universe.} 
    \label{fig: g2-Sk}
\end{figure}

\subsection{Pair Correlation Function}

We begin by considering the pair correlation function, $g_2(r)$, of the \textsc{Quijote} simulation suite. This is estimated from the array of galaxy positions using the \textsc{corrfunc} code \citep{corrfunc}, which computes the statistic in a set of bins with centers $\{r_a\}$ via:
\beq\label{eq: g2-estimator}
    \widehat g_2(r_a) = \frac{1}{N\bar\rho\,v_a}\sum_{i,j=1}^N\begin{cases} 1 & \text{if }|\vr_i-\vr_j| \text{ in bin $r_a$} \\ 0 & \text{else,}\end{cases}
\eeq
where $v_a$ is the volume of bin $a$ and $\vr_i$ is the 3D position of galaxy $i$ (accounting for periodic wrapping). In Fig.\,\ref{subfig: g2}, we display the obtained $g_2(r)$ functions, alongside corresponding results from a Poisson random sample with the same number density \resub{($\approx 1\times 10^{-4}h^{3}\mathrm{Mpc}^{-4}$)} and volume. As expected, the latter is simply unity everywhere, whilst the former shows considerable structure, and is quite different to that expected from most simple heterogeneous media \citep[e.g.,][]{To02a}. On small scales (with $r\lesssim 0.2\Mpch$, considerably less than the average galaxy separation of $\approx 8\Mpch$), $g_2(r)$ decays to zero; this is as expected, since the galaxies are of finite size and cannot overlap, enforcing some minimum separation. \footnote{Physically, galaxies \textit{can} overlap; however, they would be classed as a single object in this paradigm.}  The fact the mean galaxy (particle) separation is about an order of magnitude greater than the minimum pair separation is atypical behavior for most condensed
phase systems in which these two length scales are comparable to one another (see Refs. \cite{Han13} and \cite{To02a}). At large scales, $g_2(r)$ decays to zero slightly slower than $r^{-4}$ (in fact, $r^{-3.96}$), implying that large-scale correlations are suppressed; since the decay is between $r^{-d}$ and $r^{-(d+1)}$ for dimension $d=3$, the system is said to have \textit{quasi-long-range} correlations. This has a physical origin: the large-scale behavior of $g_2(r)$ arises from correlations in the Universe' quantum initial conditions, imprinted before cosmological inflation. Due to the dynamics of expansion (and slight breaking of time invariance in `slow-roll' inflation), these are suppressed on the largest scales. From the inset of Fig.\,\ref{subfig: g2}, we note that $g_2(r)$ has considerable structure on intermediate scales, with a prominent peak at $r\approx 100\Mpch$ sourced by acoustic oscillations fourteen billion years ago \citep{Eisenstein:1997ik}. Clearly, the correlation properties of the galactic point cloud are very different to those for most media; this arises due to the combination of Poisson-like placements of galaxies and an underlying background stochastic field from the early Universe.

\subsection{Structure Factor}

The structure factor, $\mathcal{S}(k)$, tells a similar story as the pair correlation function. This is computed by first assigning the galaxies to a grid, then computing $\tilde h(k)$ via fast Fourier transforms, here implemented using \textsc{nbodykit} \citep{nbodykit}. From Fig.\,\ref{subfig: Sk}, we observe a super-Poissonian signature on all scales, with a characteristic decline following a peak at $k\approx 0.05\hMpc$. This peak (known as the `equality peak') corresponds to a change in the Universe's expansion rate at early times, with matter starting to drive the expansion rather than radiation pressure. At larger $k$, we again see the characteristic acoustic features, here shown by oscillations in $\tilde h(k)$. On the largest scales, the power spectrum or structure factor $\mathcal{S}(k)\sim k^{0.96}$ in the infinite-wavelength limit $k\to 0$ (just visible in this plot), with a slope set by the physics of inflation, \resub{and hence because the exponent $\alpha$ in \eqref{eq:Sk-scaling} is 0.96, the Universe belongs to class III hyperuniformity, as defined in relation (\ref{eq:sigma-scaling}).  If the Universe was scale-invariant according to the Peebles-Harrison-Zeldovich spectrum with $\mathcal{S}(k)\sim k$ \cite{peebles80}, then, because $\alpha=1$, it would be hyperuniform of class II.
Of course, either scenario implies that the structure factor vanishes in the limit $k \to 0$}, the system is \textit{hyperuniform} \citep[e.g.,][]{torquato03,torquato18}. In contrast to many terrestrial media, the large-scale behavior is well understood, and can be predicted using a variety of cosmological codes; this occurs since it is an imprint of underlying dark matter physics, rather than a true pairwise interaction.

\subsection{\texorpdfstring{$\tau$}{Tau} Order Metric}



To quantify the degree of order/disorder of the the galaxies, 
we compute the metric $\tau$, which is defined by \eqref{tau}, as discussed in \S\ref{subsec: disorder}. We take $d=3$ and the \textit{unclustered} interparticle separation $D = \bar\rho^{-1/3} \approx 20\Mpch$ to be the characteristic length-scale. Here $\tau = 4.85$ for the cosmological sample, which is to be compared to $\tau=8.37\times 10^{-6}$ for the (finite-volume) Poisson realizations, close to the infinite-volume expectation of $\tau = 0$.
This result supports the well-known results that the galaxy distribution is not purely random (uncorrelated),
but instead is a correlated disordered system.
To place the magnitude of $\tau$ for the galaxies
in the context of other models of  correlated  disordered media, we compute $\tau$ 
for the random sequential addition (RSA) process, which is a time-dependent (nonequilibrium) procedure that generates disordered sphere packings in $\mathbb{R}^d$ \cite{Wi66,To06d}. Starting with
an empty but large volume in  $\mathbb{R}^d$, the RSA process is
produced by randomly, irreversibly, and sequentially 
placing nonoverlapping spheres into the volume.  If a new sphere does not overlap with any existing spheres, it will be added to the configuration; otherwise, the attempt is discarded. This procedure is repeated for ever-increasing volumes; then, an appropriate infinite-volume limit is obtained. One can stop the addition process at any time 
$t$, obtaining
RSA configurations with a range of packing fractions 
$\phi(t)$ up to the maximal `saturation' value 
$\phi(\infty)$ in the infinite-time limit, which for three dimensions is 
about 0.3812 \cite{To06d}. Using the data for pair statistics given in Ref. \cite{To06d}, we find
$\tau=6.17$ for saturated RSA packings in $\mathbb{R}^3$,
which is close in value to that of the galaxies.

\subsection{Local Number Variance}

\begin{figure}
    \begin{minipage}{0.47\textwidth}
    \centering
    \includegraphics[width=0.95\textwidth]{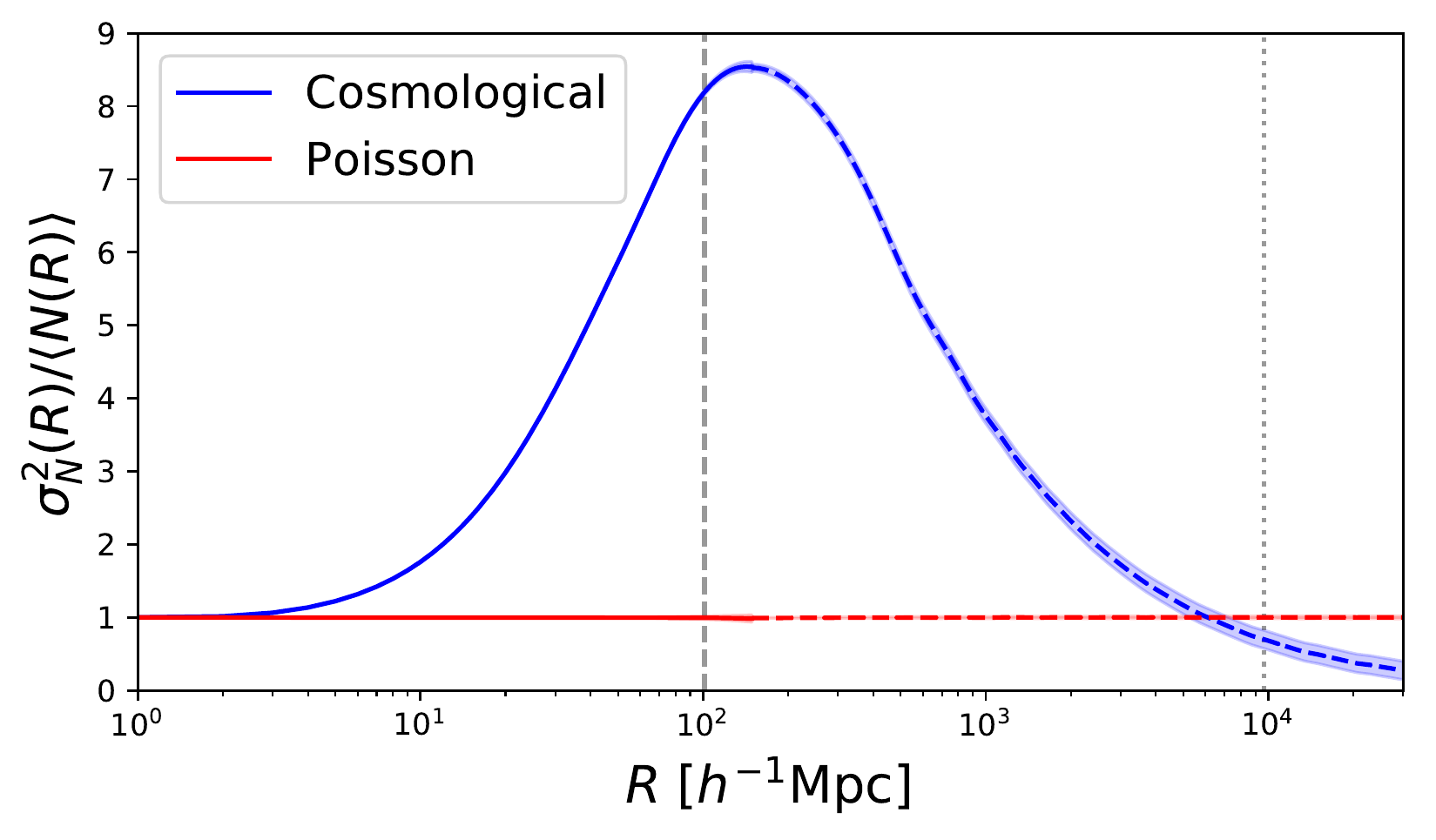}
    \caption{Local number variance for the cosmological (blue) and Poisson (red) simulations as a function of scale $R$. This is defined as the variance of the number of particles found in spheres of radius $R$, and computed directly from the correlation function shown in Fig.\,\ref{fig: g2-Sk}. The dashed lines show an extrapolation to large $R$, using the theoretical asymptotic structure factor form (with $\mathcal{S}(k)\sim k^{0.96}$). The vertical lines show two characteristic scales: the sound horizon at recombination (dashed), which sources acoustic wave in the early Universe, giving the bump in $h(r)$, and the size of the Universe as it transitioned from radiation- to matter-dominated (dotted).}
    \label{fig: sigmaN-plot}
    \end{minipage}
    \quad
    \begin{minipage}{0.47\textwidth}
    \centering
    \includegraphics[width=\textwidth]{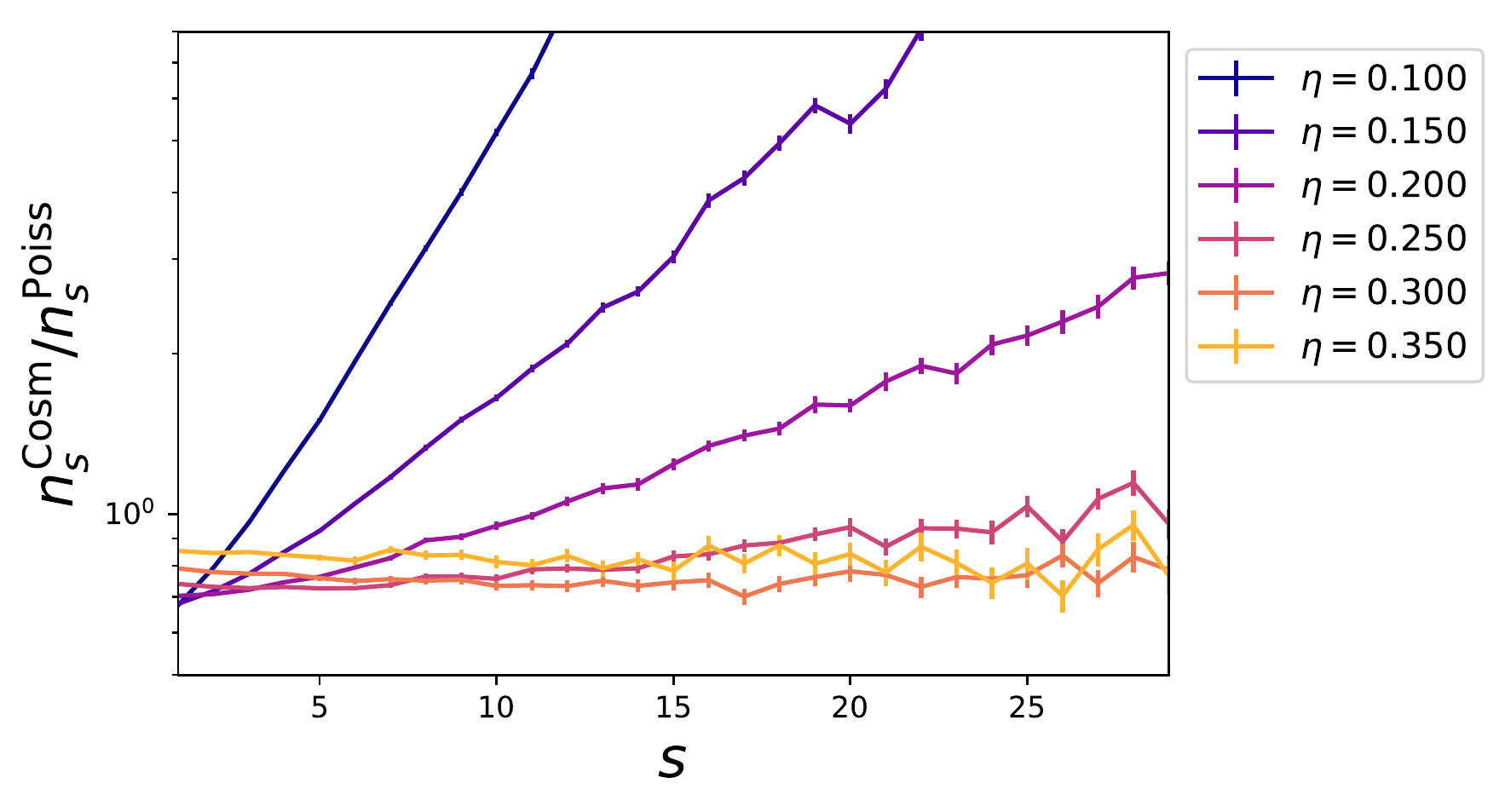}
    \caption{Comparison of the cluster size distribution in the cosmological and Poisson simulations. We plot the mean number of $s$-mers per unit particle (defined as clusters containing $s$ member) as a function of $s$, normalizing to the Poisson prediction. Results are shown for various values of the reduced density $\eta$ and we assume an aperiodic simulation volume of size $L=800\Mpch$, averaging over $100$ realizations. At large $\eta$, the two sets of simulations have a similar $n_s$-distribution (at least for small $s$), whilst at low $\eta$, the enhanced clustering in the cosmological simulations leads to significantly more $s$-mers, with the mean number of particles per cluster increasing with $\eta$ (cf.\,Fig.\,\ref{fig: cluster-sizes}).}
    \label{fig: n-s-plot}
    \end{minipage}
\end{figure}

As discussed in \S\ref{sec: statistics}, the local number variance, $\sigma_N^2(R)$, can also be computed from the measured correlation function, and provides a useful tool with which to assess the system's order. Here, this is computed from the measured $g_2(r)$ values via \eqref{eq: sigmaN-def}, and plotted in Fig.\,\ref{fig: sigmaN-plot}, alongside its extrapolation to large $R$, using the well-known large-scale limit, $\mathcal{S}(k)\sim k^{0.96}$. Notably, we find the number variance to increase faster than the Poisson case at small $R$, roughly up to the scale corresponding to the second peak in $r^2h(r)$ (arising from the imprint of acoustic oscillations from the early Universe), then fall to sub-Poisson values by scales corresponding to the peak in $\mathcal{S}(k)$.  A variance that increases
much faster than that for Poisson systems at small $R$ is unusual
for typical correlated disordered systems that have been investigated
in condensed matter physics. Of course, that the large-scale variance approaches zero indicates that the system is hyperuniform; however, these scales are difficult to measure with most cosmological surveys.

\begin{figure}
    \centering
    
    \subfloat[Void Nearest-Neighbor PDF]{%
      \includegraphics[width=0.48\textwidth]{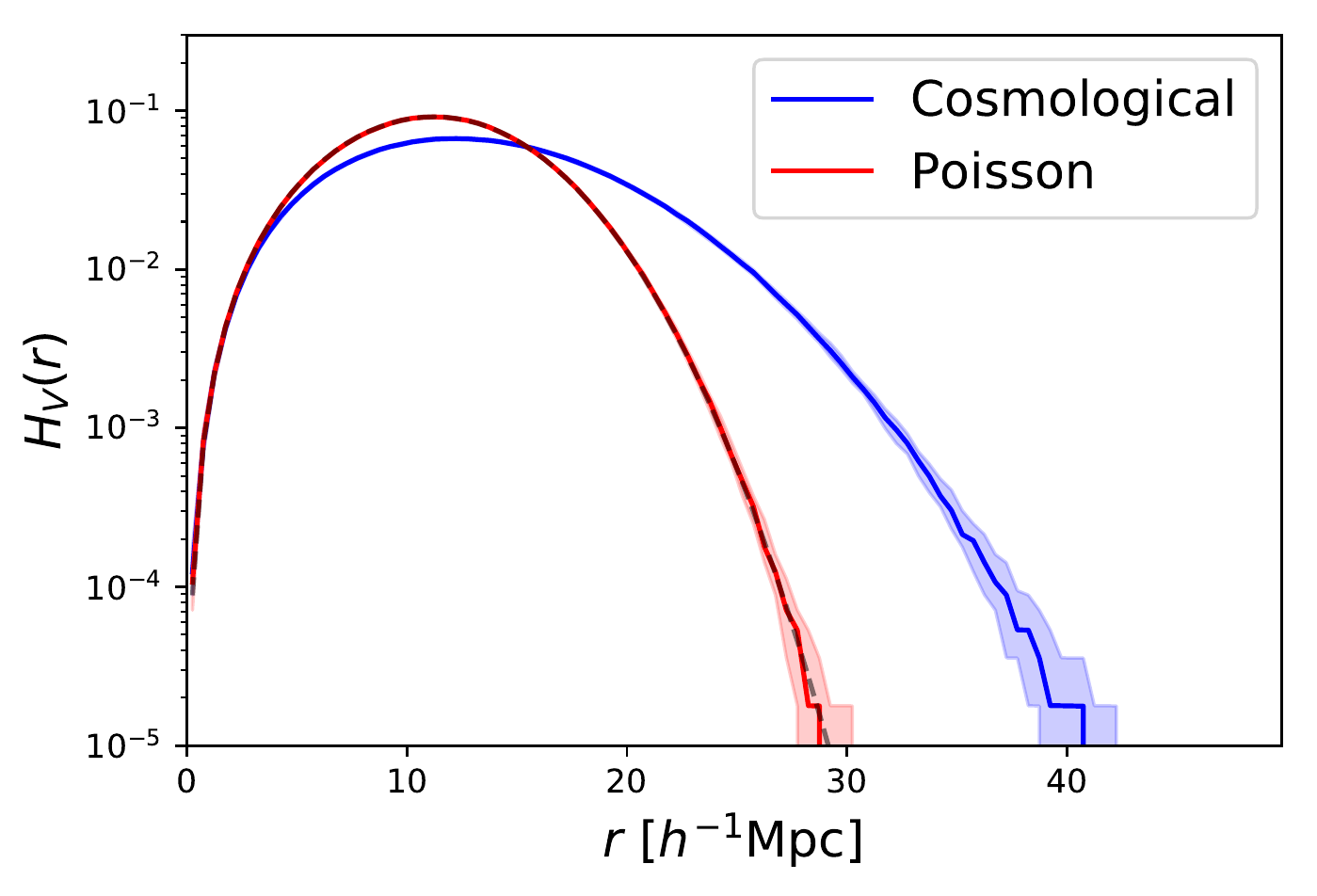}
    }
    \hfill
    \subfloat[Particle Nearest-Neighbor PDF]{%
      \includegraphics[width=0.48\textwidth]{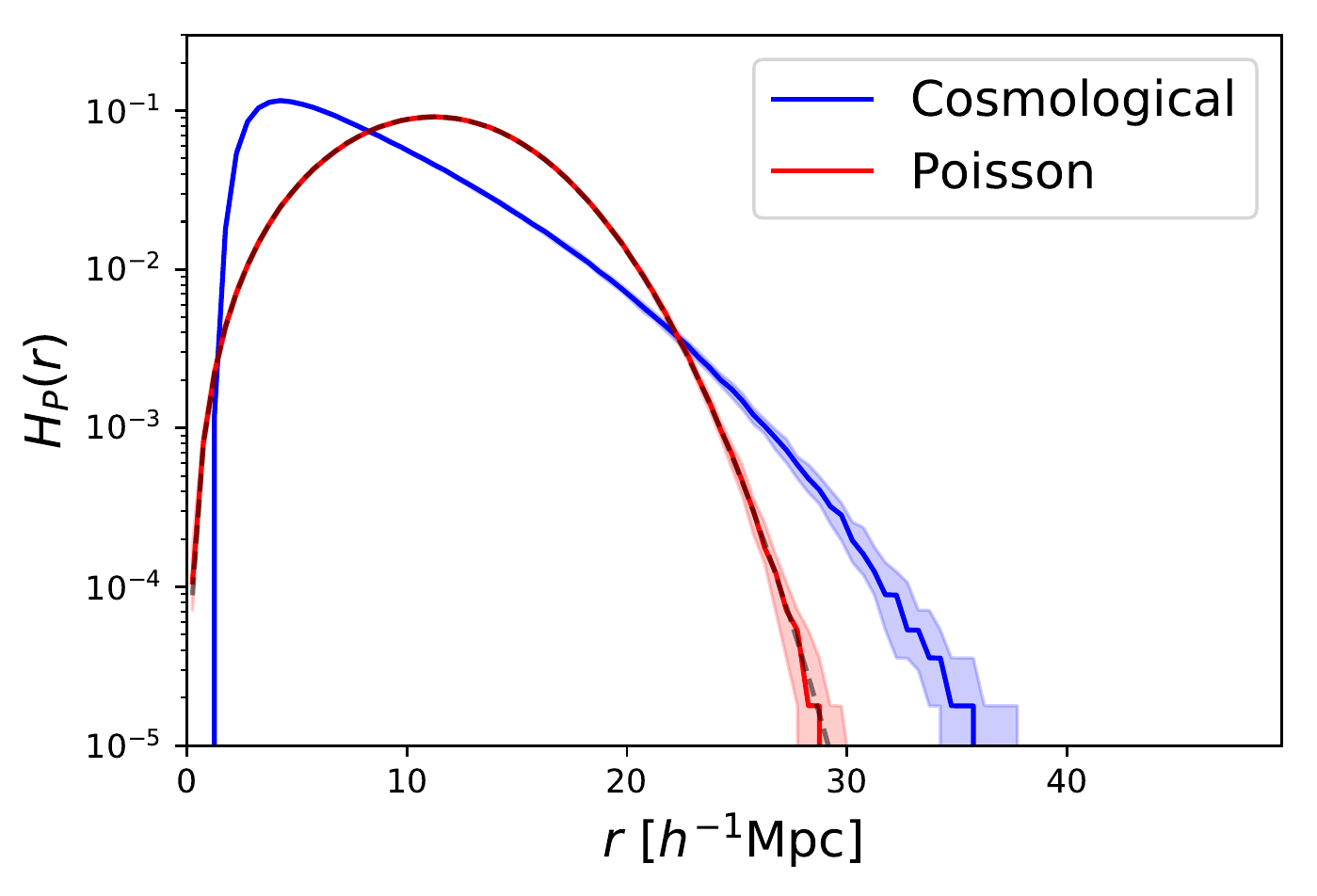}
    }
    \caption{Comparison of the void and particle nearest-neighbor probability density functions for the cosmological and Poissonian data-set, as defined in \eqref{def-Hv}\,\&\,\eqref{def-Hp}. The left panel shows the probability distribution of finding a void of radius $r$ in the cosmological (blue) and Poisson (red) datasets, averaged over 100 realizations, whilst the right panel gives the distribution function of the distance of a given particle from its nearest neighbor. For the Poisson case, the dashed lines show theory curves (defined in \S\ref{subsec: nearest-neighbor}), which are in excellent agreement with the simulations. The void distribution in the \textsc{Quijote} simulations follows the (mean-density-matched) Poisson distribution at small $r$, but has an excess of large voids (shown by much broader tails), due to the quasi-long-range correlations. In contrast, the particle distribution $H_P(r)$ differs between the cosmological and Poissonian simulations on all scales, notably with an absence of small separations (due to halo exclusion effects) and an enhancement on large scales.
    }\label{fig: vsf}
\end{figure}

\subsection{Void and Particle Nearest-Neighbor Functions} 

In Fig.\,\ref{fig: vsf}, we depict the nearest-neighbor functions of the two sets of simulations, which provide an alternative description of the system's geometrical and topological properties, as discussed in \S\ref{subsec: nearest-neighbor} \new{(see also \citep{2021MNRAS.500.5479B} for a previous discussion of the void function of galaxies, yielding similar results)}.
These are obtained from the simulations by histogramming the minimum distance between each pair of particles (for $H_P$) or a pair of particles and a Poisson random particle (for $H_V$, determining if this particle lies within a void). For the Poisson system, we find identical results for the void and particle nearest-neighbor density functions, as expected, but significant differences for the \textsc{Quijote} simulations. Whilst the cosmological case has a similar distribution of small ($r\lesssim 5\Mpch$) voids to that found in the Poisson realizations, it boasts significantly broader tail towards large $r$, and thus a somewhat larger mean void size.
Specifically, the first moment of $H_V(R)$, $\ell_V^{(1)}$, defined by (\ref{mom-V}), is equal to $14\Mpch$ ($11\Mpch$) for the cosmological (Poisson) simulations. Interestingly, the maximal void size (averaged over realizations) for the \textsc{Quijote} simulations is $42\Mpch$, which is  is almost $50\%$ larger than that for the Poisson system with a maximal void size of $30\Mpch$. Furthermore, the variance of $H_V$, defined as $\ell_V^{(2)}-\left(\ell_V^{(1)}\right)^2$ is much larger for the cosmological case: $2.7h^{-2}\mathrm{Mpc}^2$ instead of $1.5h^{-2}\mathrm{Mpc}^2$. In particular, the above results suggest that the galaxies will also boast a lower percolation threshold, foreshadowing what we describe below. 

For the particle distribution, we note that (a) the cosmological simulations have enhanced large-scale clustering, and thus a broad tail to the nearest-neighbor distance at large $r$, (b) there is a sharp cut at low $r$, with no galaxies found within a separation of $\sim 1\Mpch$. This is a consequence of `halo exclusion'; a pair of galaxies cannot be arbitrarily close, else they would be identified as a single object in the simulation code. Between these two effects, we find a reduced mean particle nearest-neighbor distribution in \textsc{Quijote}, indicating that galaxies are more likely to be found in large-scale clusters. This matches theoretical expectations. Specifically, the mean nearest-neighbor distance between particles, $\ell_P^{(1)}$, defined by (\ref{mom-P}),) is equal to $8\Mpch$ ($11\Mpch$) for the cosmological (Poisson) simulations, with a minimum distance of $1.1\Mpch$ ($0.29\Mpch$). In addition, the variance of the cosmological $H_P$, is again larger than the Poisson case, finding $2.5h^{-2}\mathrm{Mpc}^2$ instead of $1.5h^{-2}\mathrm{Mpc}^2$.


\subsection{Pair-Connectedness and Direct-Connectedness Functions}

The `astrophysical' pair-connectedness function $P_2(r)$ has not been previously studied in the literature, and is of particular interest to both cosmology and condensed matter physics. To construct this, we first take the set of $N\sim 10^5$ galaxy positions in each \textsc{Quijote} (or Poisson) simulation, and assign clusters via a `burning' algorithm (often known as `friends-of-friends' in cosmology) \citep[e.g.,][]{Davis:1985rj,stauffer2018introduction}, here using the \textsc{nbodykit} implementation \citep{nbodykit}. This finds sets of points for which each member is connected to each other member via a path through the clustered phase formed of spheres of radius $D(\eta)=\left[6\eta/\pi\bar\rho\right]^{1/3}$ around each point, where $\eta$ is the reduced density. Given the set of particles and cluster memberships (visualized in Fig.\,\ref{fig: density}), we compute the pair-connectedness function in bins with centers $\{r_a\}$ via
\beq
    \widehat P_2(r_a,\eta) = \frac{1}{N\bar\rho v_a}\sum_{i,j=1}^N\begin{cases} 1 & \text{if }|\vr_i-\vr_j| \text{ in bin $r_a$} \textbf{ and } \text{$i$, $j$ in same cluster} \\ 0 & \text{else.}\end{cases},
\eeq
analogous to \eqref{eq: g2-estimator}. This is achieved using a custom modification of the \textsc{corrfunc} code \citep{corrfunc}, which accepts pairs only if they have the same cluster index.

\begin{figure}
    \centering
    \includegraphics[width=\textwidth]{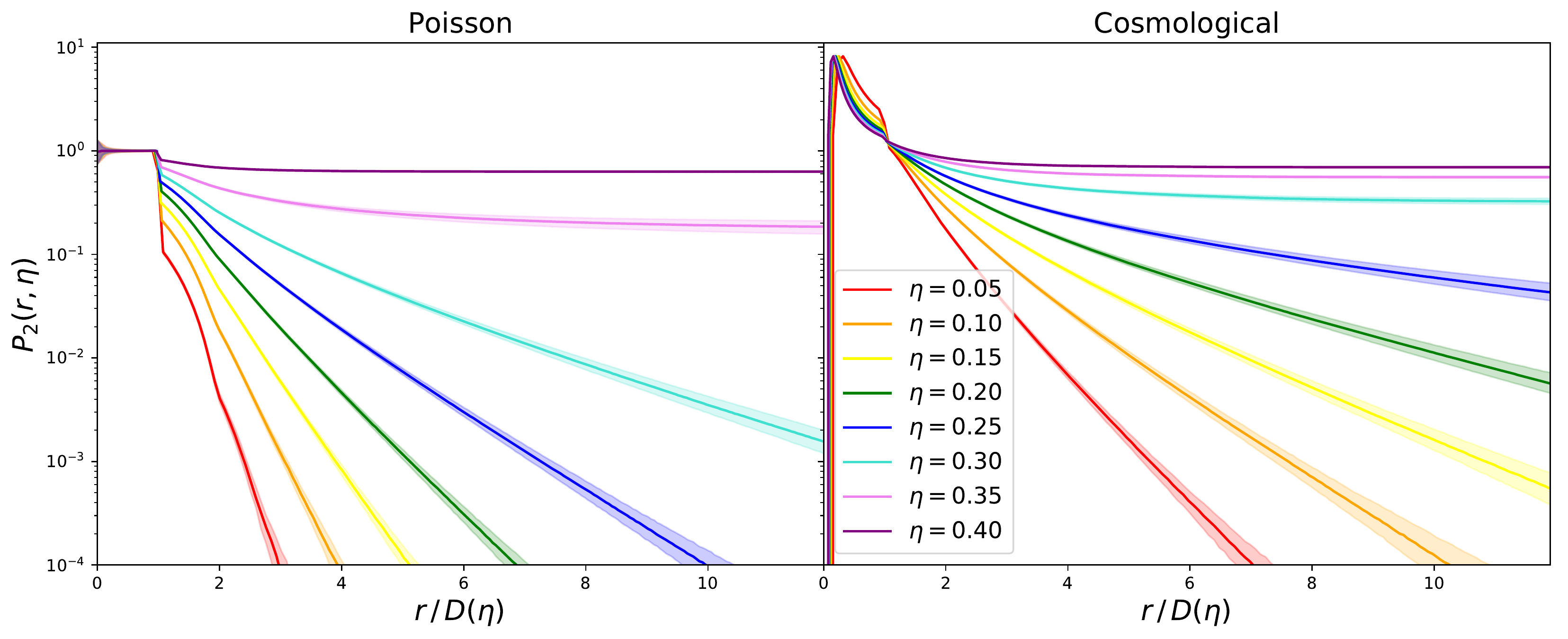}
    \caption{Pair-connectedness function, $P_2(r,\eta)$, for the Poisson (left) and cosmological (right) simulations. Results are shown for a variety of reduced densities $\eta$, corresponding to clustering distances $D(\eta)$ in the range $[10,20]\Mpch$, and shaded regions show the $1\sigma$ deviations expected from statistical fluctuations. The \textsc{Quijote} simulations show significantly enhanced correlations on large scales, due to the underlying correlations of matter imprinted in the early Universe. This additionally suggests that the galaxy sample will percolate at lower $\eta$: this will be explored in \S\ref{sec: percolation}.}
    \label{fig: P2-plot}
\end{figure}

Figure\,\ref{fig: P2-plot} displays the pair correlation functions from the \textsc{Quijote} simulations alongside the more familiar Poisson case. The latter match our expectations: $P_2(r) = 1$ for $r<D$ (since all particles with this separation must be in the same cluster), and $P_2(r)$ falls sharply with $r$ for $r>D$ (due to an absence of large-scale clusters), with an enhanced decline at low $\eta$. For large $\eta$, the volume integral of $P_2(r)$ appears to diverges (at least in the infinite volume limit), indicating percolation.\footnote{Note that the simulations are computed in periodic boxes, which are known to be suboptimal for computing $P_2(r)$ on the largest scales \citep{lee-torquato90}. This will be addressed in \S\ref{sec: percolation} in the context of finite scaling analyses.} For the cosmological simulations, we firstly note that $P_2(r) = g_2(r)$ for $r<D$, as expected. At larger $r$, we find that $P_2^{\rm sim}(r)>P_2^{\rm Poiss}(r)$ for all choices of $\eta$, indicating that our galaxy catalogs contain more long-range correlations than a Poisson random field of the same density, and suggesting that the system will also percolate quicker. In the large $\eta$ limit (\textit{i.e.}\ above percolation), $P_2(r) \to g_2(r)$, since all points belong to the same cluster. The cosmological utility of $P_2(r,\eta)$ will be discussed in \S\ref{sec: pair-connectedness}.

\begin{figure}
    \centering
    \includegraphics[width=\textwidth]{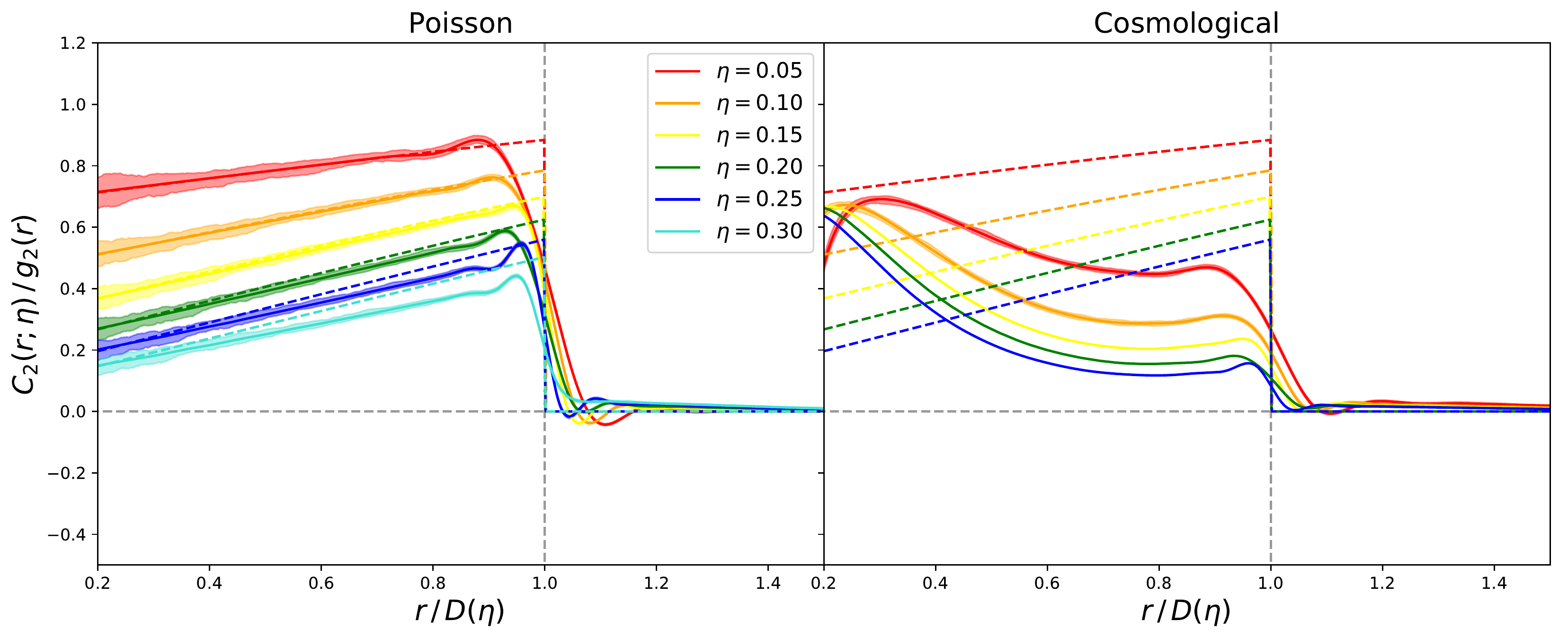}
    \caption{Direct-connectedness function, $C_2(r,\eta)$, for the Poisson (left) and cosmological (right) simulations. This follows Fig.\,\ref{fig: P2-plot}, but focusses on smaller scales, and additionally includes analytic predictions from the Percus-Yevick model. We additionally normalize all quantities by $g_2(r)$, and exclude values of $\eta$ for which $C_2(r,\eta)$ is not well-behaved (beyond the percolation threshold).}
    \label{fig: C2-plot}
\end{figure}

The direct-connectedness function also plays an important role in the analysis of connected systems, in part due to its appearance in the Ornstein-Zernike equation \eqref{eq: OZ-Poisson}. Given $P_2(r)$, this can be computed using \eqref{eq: P2-C2-relation}, performing the Fourier transforms numerically via the FFTLog prescription \citep{hamilton2000}. Figure\,\ref{fig: C2-plot} shows $C_2(r,\eta)$ for both the cosmological and Poisson simulations, alongside the analytic `Percus-Yevick' model, which solves the OZ equation by asserting that $C_2(r>D,\eta) = 0$ and $P_2(r<D,\eta)=1$ \citep{percus-yevick}.\footnote{This is computed for point objects by using the correspondence with the known (cubic) form for hard spheres via $C_2^{\rm PY, Poiss}(r,\eta) = -C_2^{\rm PY, hard\,sphere}(r,-\eta)$ \citep{Stell_1984}.} For the Poisson case, we find good agreement between theory and simulations for small $\eta$ (far from the percolation threshold of $\eta_c\approx 0.34$), particularly away from the boundary at $r = D(\eta)$. We observe very little power from the region with $r>D(\eta)$, since most intra-cluster path-ways with $r>D(\eta)$ contain at least one node, and thus do not contribute to $C_2(r)$. The cosmological simulations show a very different behavior, with two peaks observed, with one in similar location to the Poisson system and one at smaller $r$. This statistic represents the complexities of the clustering on smaller scales than that typically seen in $P_2$ (with $D(\eta)\sim 10\Mpch$), and the differences arise primarily due to small-scale physics, such as the restriction that galaxies cannot be arbitrarily close together. We also note that $C(r,\eta)$ was found to be ill-behaved for $\eta\gtrsim 0.3$ (due to $P_2(r,\eta)$ not being square integrable), indicating that the cosmological simulations have percolated by around this value of $\eta$ (cf.\,\S\ref{sec: percolation}). In practice, we expect the percolation threshold to depend on the peculiarities of the galaxy sample in question: this will be discussed further below.

\section{Percolation and Fractal Dimensions}\label{sec: percolation}







We now turn to the issue of \textit{percolation}, following the discussion in \S\ref{subsec: percolation-stats}. As noted earlier, determining
the mean cluster size $S(\eta)$ in a system as a function of $\eta$ is a useful way in which to test whether a system has reached percolation. We utilize
the representation of $S(\eta)$ in terms of $s$-mer cluster statistic, $n_s$,
as defined by (\ref{S-ns}). In principle, we expect $S(\eta)\to\infty$ as $\eta\to\eta_c$; in practice, $S(\eta)\leq N$, where $N$ is the total number of particles in the box. To account for this, it is useful to analyze a number of different configurations with different boxsizes, $L$ (and thereby $N\equiv\bar\rho L^3$). Here, we construct (aperiodic) subboxes from the \textsc{Quijote} simulations, with $L$ in the range $[400,800]\Mpch$ (noting that the majority of our analyses are restricted to $r<200\Mpch$), and construct analogous Poisson realizations for each. To examine percolation at each choice of boxsize, we generate clusters for various values of $\eta$ by varying the sphere radius $D(\eta)$, and utilizing burning (`friends-of-friends') algorithms, as described above.

\begin{figure}
    \centering
    \includegraphics[width=\textwidth]{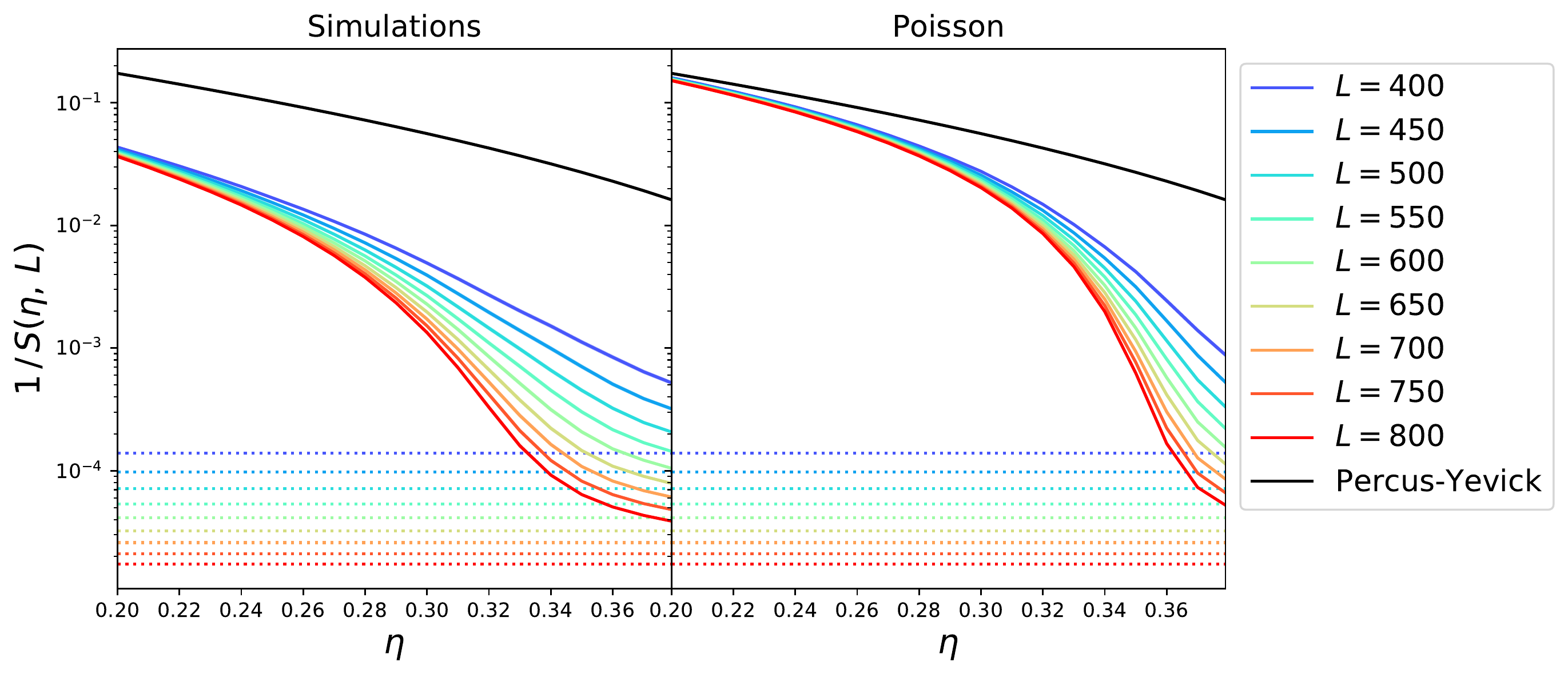}
    \caption{Mean number of particles per cluster, $S(\eta,L)$, as a function of the reduced density $\eta$ and (cubic) simulation boxsize $L$. We show results for both cosmological data and Poisson realizations, plotting the reciprocal $1/S(\eta,L)$ averaged over 1000 realizations. Note that $S(\eta)$ is bounded by the number of particles in the data-set ($N$), shown by dotted lines. We additionally show the Percus-Yevick prediction in black, which is inaccurate for all but the smallest $\eta$.}
    \label{fig: cluster-sizes}
\end{figure}

Figure\,\ref{fig: cluster-sizes} shows the mean cluster size for the two data-sets as a function of $L$ and $\eta$. In both cases, we observe that $S(\eta,L)$ begins to approach its asymptotic limit as $\eta$ increases, and, moreover, the limit is approached faster as the boxsize increases. Extrapolating the Poissonian results to large $L$, the percolation threshold (whence $S(\eta,L)\approx N$) appears to be around $\eta_c = 0.35$, matching previous studies \citep{lee-torquato88}. For the cosmological simulations, we find a generally slower approach to $\eta_c$ (corresponding to a different critical exponent), and additionally a lower percolation threshold, around $\eta_c = 0.28$ for $L\to\infty$. As above, this arises since the galaxy sample contains a stochastic background inhomogeneity, leading to various areas being super- or sub-Poisson populated in a correlated manner. The difference is evident even low $\eta$: for $\eta=0.1$, clusters in the full-volume cosmological simulation contain an average of $\approx$ 15 particles, whilst those in the Poisson realizations contain $\approx 2$ only.

It is further instructive to consider the size distribution of clusters, via the average number of $s$-mers, $n_s$, as defined in \eqref{eq: S-eta-def}. This is shown in Fig.\,\ref{fig: n-s-plot} for a suite of cosmological and Poissonian boxes at $L=800$ with a variety of values of the reduced density $\eta$. At low $\eta$, we find that the ratio of cosmological and Poissonian simulations is a strongly increasing function of $s$, with the largest slopes seen for small reduced densities. In this limit, the system is far from percolation, thus large clusters are rare in both systems. The enhanced correlations in the galaxy distribution seen in the cosmological case increase the probability of an $s$-mer forming (at fixed $s$), giving this stark difference in behavior. As $\eta$ approaches the percolation threshold, the $n_s$-ratio becomes roughly constant with $s$; this indicates that the additional galaxy correlations impacts only the largest $s$-mers, as we are dominated by the clustering signal imprinted by the circumscribed spheres, rather than any intrinsic effects.

\begin{figure}
    \centering
    \includegraphics[width=\textwidth]{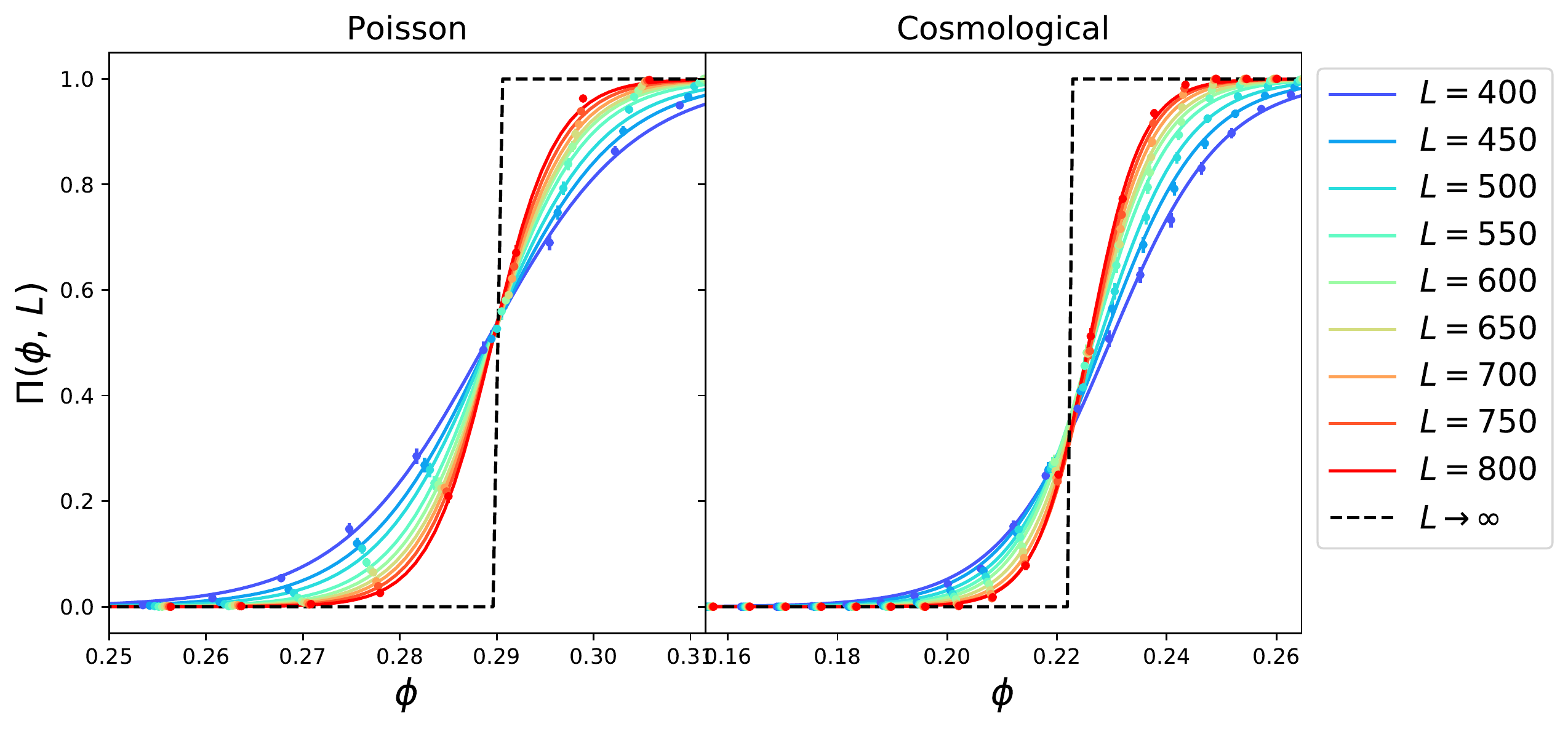}
    \caption{Percolation probability, $\Pi(\phi,L)$, obtained from 1000 \textsc{Quijote} and Poisson simulations, for various non-periodic boxsizes, $L$, and volume filling fractions, $\phi$. This is defined as the fraction of realizations containing a sample spanning cluster, and asymptotes to a Heaviside function centered at the percolation threshold for $L\to\infty$ (shown as a dotted line). Points represent the values computed from simulations, whilst the solid lines show a fit using the sigmoid function of \eqref{eq: sample-spanning-prob-sigmoid}. Using a finite scaling analysis, we find the percolation thresholds of $\eta_c = 0.252$ ($0.343$) for the \textsc{Quijote} (Poisson) simulations, though we caution that the cosmological result depends on the sample density.}
    \label{fig: finite-scaling}
\end{figure}

To measure the percolation threshold of the cosmological simulations in a robust fashion, we perform a \textit{finite scaling analysis}, following the approach of \citep{xu19}, originally formulated in \citep{fisher61}. In essence, this computes the percolation probability ($\Pi(\eta,L)$, defined as the fraction of realizations containing a cluster for which the circumscribed spheres overlap with both the top and bottom of the box) for the simulations at various values of $L$ and $\eta$ and extrapolate using asymptotic scaling relations to find the $L\to\infty$ limit. Figure\,\ref{fig: finite-scaling} shows the obtained percolation probability distribution for both sets of simulations as a function of the volume filling fraction $\phi$. This is computed numerically for each simulation from the probability that a randomly chosen point within the box is within a distance $D/2$ from the nearest particle, \textit{i.e.}\ whether it is within the sphere phase; for the Poisson case, this is asymptotically equal to $1-e^{-\eta}$. The behavior seen in Fig.\,\ref{fig: finite-scaling} is qualitatively similar for the Poisson and cosmological system: the percolation probability is small for low $\phi$ (whence the typical extent of the cluster is far below $L$), and asymptotes to unity at large $\phi$. As the boxsize increases, the transition becomes sharper, and asymptotes to a Heaviside function in the $L\to\infty$ limit. It is also clear that the cosmological simulations percolate at smaller values of $\eta$ than the Poisson realizations; this is as expected, and indicates their enhanced clustering due to background inhomogeneities.

To extract the percolation thresholds from $\Pi(\phi,L)$, we fit the data to the phenomenological sigmoid model of \citep{xu19}, as shown in Fig.\,\ref{fig: finite-scaling}:
\beq\label{eq: sample-spanning-prob-sigmoid}
	\Pi(\phi,L) \approx \frac{1}{2}\left[1+\mathrm{tanh}\left(\frac{\phi-\phi_c(L)}{\Delta(L)}\right)\right]
\eeq
where $\phi_c(L)$ and $\Delta(L)$ are the percolation volume fraction and width at boxsize $L$ and $\phi$ is the volume filling fraction obtained as described above. Asymptotically, $\Delta(L)\sim L^{-1/\nu}$ and $\phi_c(L)-\phi_c \sim L^{-1/\nu}$ for critical exponent \citep[e.g.,][]{To02a}; by fitting for $\nu$ from the obtained values of $\Delta(L)$, we can thus obtain $\phi_c\equiv \lim_{L\to\infty}\phi_c(L)$. Here, we find a critical exponent of $\nu = 0.85\pm0.03$ ($0.88\pm0.03$) for the \textsc{Quijote} (Poisson) simulations, with a corresponding percolation threshold of $\phi_c = 0.223$ ($0.290$) or $\eta_c = 0.252$ ($0.343$), each with a statistical error around $0.001$. The Poissonian case matches standard results \citep[e.g.,][]{To02a}, and, as foreshadowed in Figs.\,\ref{fig: P2-plot}\,\&\,\ref{fig: cluster-sizes}, the cosmological system percolates at lower densities, due to the additional clustering signature imprinted by early-Universe and galaxy formation physics. In addition, the fact that the two sets of simulations appear to share the same critical exponent $\nu$ suggests that they belong to the same \textit{universality class}, as do other correlated disordered
systems \cite{Le90a}.

It is important to note that the percolation thresholds found herein are not a universal property of galaxy distributions; rather, they depend on the galaxy sample in question. To explore this, we have repeated the analysis using a galaxy sample with half the density of the fiducial sample, and another including dark matter halos (\textit{i.e.}\ galaxies) down to half of the aforementioned minimum size. For the former case (with $\bar\rho\sim 0.5\times 10^{-4}h^{3}\mathrm{Mpc}^{-3}$, we find that the percolation threshold for the cosmological sample increases to $\eta_c = 0.285$, whilst remains the same for the density-matched Poissonian sample (as expected). This can be rationalized by noting that the galaxies roughly follow Poisson statistics above a stochastic background, caused by the matter density; if $\bar\rho$ is reduced, the Poisson part of the stochasticity becomes more dominant, thus $\eta_c$ tends towards its Poisson limit. In the second scenario, we find $\eta_c = 0.207$, significantly lower than the fiducial analysis. In this case, we have both a sample of almost twice greater density, and one that is more biased with respect to the continuous dark matter density (such that $\hat\rho_{\rm galaxy}/\hat\rho_{\rm dark\ matter}$ is larger, smoothed on sufficiently large scales). In both cases, however, we find a similar critical exponent, ($0.82\pm0.03$) to the above.

Finally, we consider the effective fractal dimension of the system, $d_F$. As discussed in \S\ref{sec: statistics}, this may be computed from the dependence of the sample spanning cluster mass, $M(\eta,L)$ (\textit{i.e.}\ its number of constituent particles) on the simulation boxsize $L$ at the percolation threshold $\eta_c$ \eqref{eq: M-L-scaling}. To explore this, we repeat the above analysis for the fiducial sample, computing the mass of the sample spanning cluster (when it exists) for five boxsizes in the range $[700,800]\Mpch$ and five reduced densities in the range $\eta_c\pm0.1$. For the Poisson system, fitting for the relationship $M(\eta,L)\sim L^{d}$ and interpolating to $\eta_c$ gives $d_F\equiv d(\eta_c)\approx2.40\pm0.07$, matching that predicted from theory \citep[e.g.,][]{To02a}. For the galaxy sample, we find $d_F = 2.36\pm0.08$, which is consistent with the Poisson realizations, even though the percolation threshold differs. This is an important result: the cosmological sample lies in the same universality class as simple Poisson realizations, for the range of scales considered: $r\sim 500-1000\Mpch$. This is broadly consistent with previous results on smaller scales; \resub{\citep{baryshev05,gabrielli05,1988ApJ...335L..43B} describes a variety of methods to ascertain the effective fractal index}, with galaxy counts yielding $d_F = 2.2\pm 0.2$ on $\lesssim 10\Mpch$ scales, and correlation functions finding the same on $\lesssim100\Mpch$ scales.





\section{The Pair-Connectedness Function as a Cosmological Descriptor}\label{sec: pair-connectedness}

\subsection{Background}

A crucial problem in modern-day cosmology is the extraction of physical parameters from observed statistics, such as the distribution of galaxies. In the standard paradigm (dubbed $\nu\Lambda$CDM), six parameters are of relevance: (1) the Universe's current expansion rate, $H_0$, (2) the density of baryonic matter, $\omega_b$, (3) the combined density of dark matter and baryonic matter, $\Omega_m$, (4) the amplitude of clustering in the Universe, $\sigma_8$, (5) the slope of the primordial power spectrum (\textit{i.e.}\ structure factor), $n_s$, (6) the sum of the neutrino masses, $\sum m_\nu$. Whilst $n_s$ and $\Omega_b$ are well constrained by observations of the cosmic microwave background \citep[e.g.,][]{planck18}, the remaining parameters are a key target for upcoming galaxy surveys. Traditionally, they are constrained \new{through summary statistics such as the two-particle and three-particle correlation functions. Such an analysis proceeds by the fitting measured statistics} to analytic models depending on the above physical parameters \citep[e.g.][]{philcox22}. 

In this section, we consider \resubtwo{the utility of alternative statistics (described above) in this effort. Three metrics by which we judge a statistic to be useful are: (a) ease of computation, (b) dimensionality, (c) information content.}
Here, we will principally concentrate on the pair-connectedness function, $P_2(r)$, \resubtwo{since this has not been previously been used in cosmology,} unlike void probability \new{or nearest-neighbor} functions \citep{2022MNRAS.514.3828W}. \resubtwo{As shown above, the statistic is simple and fast to measure from the data, and has a low-dimensional form, satisfying two of the above criteria. We now proceed to quantify its ability to constrain cosmological parameters.}

\new{An alternative approach \resubtwo{to the above prescription} is to model the entire galaxy distribution directly (without compressing to statistics such as the correlation functions), either with perturbative methods \citep[e.g.,][]{Cabass:2020jqo,Cabass:2019lqx,Schmidt:2020viy,Schmittfull:2018yuk} or machine learning approaches \citep[e.g.,][]{Modi:2021acq,Seljak:2017rmr,Dai:2022dso,Jamieson:2022daw,Jamieson:2022lqc,AlvesdeOliveira:2020yix}. In principle, this approach enables one to obtain optimal constraints on all parameters of interest, though is non-trivial to implement in practice, due to the huge dimensionality of the galaxy distribution and the necessity to run a large number of expensive simulations.} 





\subsection{Quantifying Information Content}

Standard cosmological analyses proceed by measuring a set of statistics from a dataset, then comparing them to accurate physical models depending on cosmological parameters, including those discussed above. If the noise properties of the statistics are known (for example, if we assert that the distribution of $g_2$ is a multivariate Gaussian), this comparison can be used to place constraints on the underlying parameters via Bayes theorem. A useful estimate of the constraining power of some statistic $X$ (e.g., $g_2$) can be obtained using a \textit{Fisher matrix} \citep{fisher22}, defined as
\beq\label{eq: fisher}
	F^X_{\alpha\beta} = \left(\frac{\rmd X}{\rmd\theta_\alpha}\right)^T\mathsf{C}_X^{-1}\left(\frac{\rmd X}{\rmd\theta_\beta}\right), 
\eeq
where $\{\theta_\alpha\}$ are the set of cosmological parameters of interest, and $\mathsf{C}$ is the covariance matrix of $X$ (treated as a vector), \textit{i.e.}\,$\mathsf{C}_X = \mathbb{E}\left[X\,X^T\right]$, averaging over realizations of the underlying microscopic density at fixed $\theta$. According to the Cram\'{e}r-Rao theorem, $\left(F^{-1,X}\right)_{\alpha\alpha}$ gives the best possible constraint on $\theta_\alpha$ from a measurement $X$, \textit{i.e.} $\mathrm{var}(\theta_\alpha) \geq \left(F^{-1,X}\right)_{\alpha\alpha}$.\footnote{This limit is saturated if $X$ obeys Gaussian statistics, \textit{i.e.}\ $\hat X\sim \mathcal{N}\left(X(\theta),\mathsf{C}\right)$.} To assess the utility of statistics such as $P_2$ and $g_2$, we need simply compute the covariance matrix and the parameter derivatives appearing in \eqref{eq: fisher}, both of which can be done using a set of simulations. Explicitly, given a set of $n$ realizations $\hat X^{(i)}$ with varying initial conditions, the two can be computed via
\beq\label{eq: cov-deriv-def}
    \widehat{\mathsf{C}}_X &=& \frac{1}{n-1}\sum_{i=1}^n\left(\hat X^{(i)}-\overline X\right)\left(\hat X^{(i)}-\overline X\right)^T,\\\nonumber
    \widehat{\frac{\rmd X}{\rmd\theta_\alpha}} &=& \frac{1}{n}\sum_{i=1}^n\frac{\hat X^{(i)}(\theta_\alpha+\delta\theta_\alpha)-\hat X^{(i)}(\theta_\alpha-\delta\theta_\alpha)}{2\,\delta\theta_\alpha},
\eeq
using finite-difference for the parameter derivatives, and denoting $\bar X = (1/n)\sum_{i=1}^n\hat X^{(i)}$ \citep[e.g.,][]{quijote}.

Whilst the Fisher forecast is appealing in its simplicity, it is not without limitations. Firstly, it gives accurate bounds on cosmological properties only if the statistics are Gaussian distributed, which can break down in the case of large correlations between bins, and the parameter posterior is Gaussian, which fails for non-negative parameters, for example. Secondly, a large number of simulations may be required to compute the quantities in \eqref{eq: cov-deriv-def}, and, if too few are used, the constraining power of a given statistic will be overestimated.\footnote{This occurs since noise in the parameter derivatives add a positive definite contribution to the Fisher matrix, and thus reduce the size of the inverted matrix, \textit{i.e.} the output parameter variances.} An alternative approach is to use \textit{simulation-based inference} (also known as `likelihood-free analysis') \citep[e.g.,][]{Papamakarios16,Alsing19,cranmer20}. In essence, this draws a set of cosmological parameters $\theta$ from some input prior, computes a realization $\hat X(\theta)$ for each, and compares a `true' data-set to the empirical distribution from the simulations. This does not make assumptions on the statistics' noise properties, and, in the case of too few simulations, will only \textit{underestimate} the cosmological utility.

Here, we examine the constraining power of various summary statistics using both the Fisher matrix formalism (which has become commonplace in cosmology) and simulation based inference (which is far less common, though more accurate). In particular, we consider the cosmological parameters $H_0$, $\Omega_m$, $\sigma_8$ and $\sum m_\nu$ and the following descriptors: $g_2$, $g_3$, $P_2$, and $H_V$, \resub{all of which can be defined for discrete point clouds, such as the galaxy density used in this work}. \resub{For the pair-connectedness function, we fix the reduced density to $\eta\equiv\bar\rho\pi D^3/6=0.2$, which is a useful balance between the uninformative case ($\eta=0$) and the percolated limit discussed in \S\ref{sec: percolation}, though we note that other choices may yield somewhat different results. Additional} statistics could be straightforwardly added, though we caution that descriptors such as the number variance are fully described by $g_2$, and will thus not add additional information.
Using the \textsc{fastpm} code \citep{Feng:2016yqz}, we run $n=512$ \textsc{Quijote}-like dark-matter simulations with the following fiducial parameters: $\{H_0 = 68\,\mathrm{km}\,\mathrm{s}^{-1}\mathrm{Mpc}^{-1}, \Omega_m = 0.31, \omega_b = 0.0227, n_s = 0.96, \sigma_8 = 0.8178, \sum m_\nu = 0\,\mathrm{eV}\}$, and compute the four statistics for each realization. 

\begin{figure}
    \centering
    \includegraphics[width=0.5\textwidth]{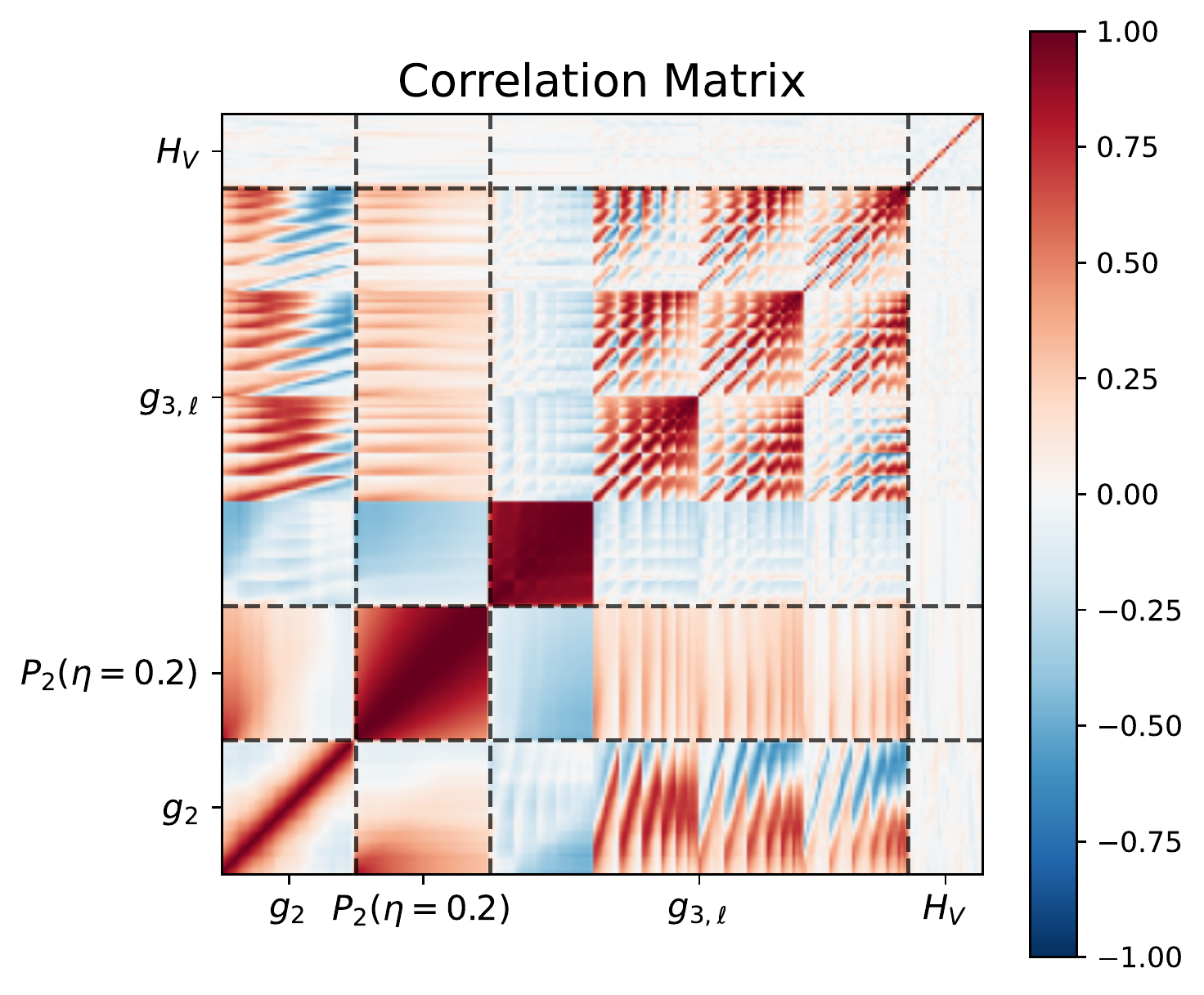}
    \caption{Correlation matrix of the pair correlation function ($g_2$), the pair-connectedness function ($P_2$, with reduced density $\eta=0.2$), the three-particle correlation function ($g_3$) and the void nearest-neighbor function ($H_V$) obtained from $512$ cosmological simulations run with a fiducial set of cosmological parameters. The correlation matrix is defined as the covariance matrix normalized by its leading diagonal, \textit{i.e.} $C_{ij}/\sqrt{C_{ii}C_{jj}}$, with all values lying in the range $[-1,1]$. The labels show the statistic of interest, which are demarcated by the dotted lines. In each statistic, the bins are ordered from small-scale (bottom left) to large-scale (top-right). For $g_3$, we use four Legendre multipoles, $g_{3,\ell}$ with $\ell\in\{0,1,2,3\}$. Notably, the individual bins of $g_{3,0}$ and $P_2$ are highly correlated, and there are a number of non-trivial correlations between observables, though few with $H_V$.}
    \label{fig: corr-matrix}
\end{figure}

In all cases, we consider only scales above $20\Mpch$ (where the simulations are accurate, \resubtwo{given the mean pairwise separation of $\approx 8\Mpch$}), 
and choose the radial bin sizes to keep the dimensionality fixed to $\mathcal{O}(100)$ elements. $g_2$, $P_2$, and $H_V$ are computed as before, 
with $g_3$ computed using the approach of \citep{Slepian:2015qza}, involving a decomposition into a Legendre multipole basis, with components $g_{3,\ell}(r_1,r_2)$ for $\ell\in\{0,1,2,3\}$. \resubtwo{In accordance with \S\ref{sec: galaxies}, we compute this statistic in configuration-space (rather than as a Fourier-space bispectrum), which obviates the need to grid the particles.} These simulations are used to compute the covariance matrix $\mathsf{C}$ of the statistics using \eqref{eq: cov-deriv-def},\footnote{Initial testing demonstrated that this is a sufficient number of simulations to estimate $\mathsf{C}$ and its inverse robustly, after including the correction factor of \citep{hartlap07}.} the structure of which is visualized in Fig.\,\ref{fig: corr-matrix}. We find significant correlations both within and between a number of statistics. In particular, the individual bins of $g_3$ and $P_2$ are highly correlated, indicating that their noise properties may not be Gaussian. In contrast, the void nearest-neighbor function has an almost diagonal correlation matrix, and is seen to be largely independent from other statistics. This suggests that it can add significant information compared to analyses using $g_2$-alone.

The other ingredient required for Fisher forecasting is the set of parameter derivatives. These are computed using \eqref{eq: cov-deriv-def}, with $n=512$ simulations (again computed using \textsc{fastpm}, with a total cost of $\sim 10^4$ CPU-hours), utilizing finite difference in each of the eight sets of parameters. For the neutrino mass, we have the bound $\sum m_\nu>0$, thus we instead utilize one-sided derivatives, following \citep{quijote}, and using the method of \citep{Bayer:2021kwg} to emulate the effects of massive neutrinos by modifying the initial conditions. Following this, we compute the Fisher matrix via \eqref{eq: fisher} for various combinations of statistics. We caution that this result appears to retain some dependence on $n$ due to residual noise in the parameter derivatives. This will lead to the constraints being artificially tightened somewhat; however, it is computationally impractical to increase $n$ by a significant amount. 

For the simulation-based inference (hereafter SBI), we utilize a set of $8192$ galaxy simulations computed using \textsc{fastpm} with the method of \citep{Bayer:2021kwg} at random locations in parameter space, according to the flat priors: $H_0\in[0,100]\mathrm{km}\,\mathrm{s}^{-1}\mathrm{Mpc}^{-1}$, $\Omega_m\in[0,0.3]$, $\sigma_8\in[0.4,1.2]$, $\sum m_\nu\in[0,4]\mathrm{eV}$.\footnote{The neutrino mass limit is significantly weaker than the bound from the latest probes \citep{planck18}, but is appropriate given the small volume of the simulations.} Summary statistics for each are computed as before, and fed into the \textsc{sbi} code, which uses neural networks (via the `Sequential Neural Posterior Estimation' method) to compute the parameter posterior, given a `true' observation, taken from the mean of the fiducial simulations discussed above.

\begin{figure}
    \centering
    \includegraphics[width=0.45\textwidth]{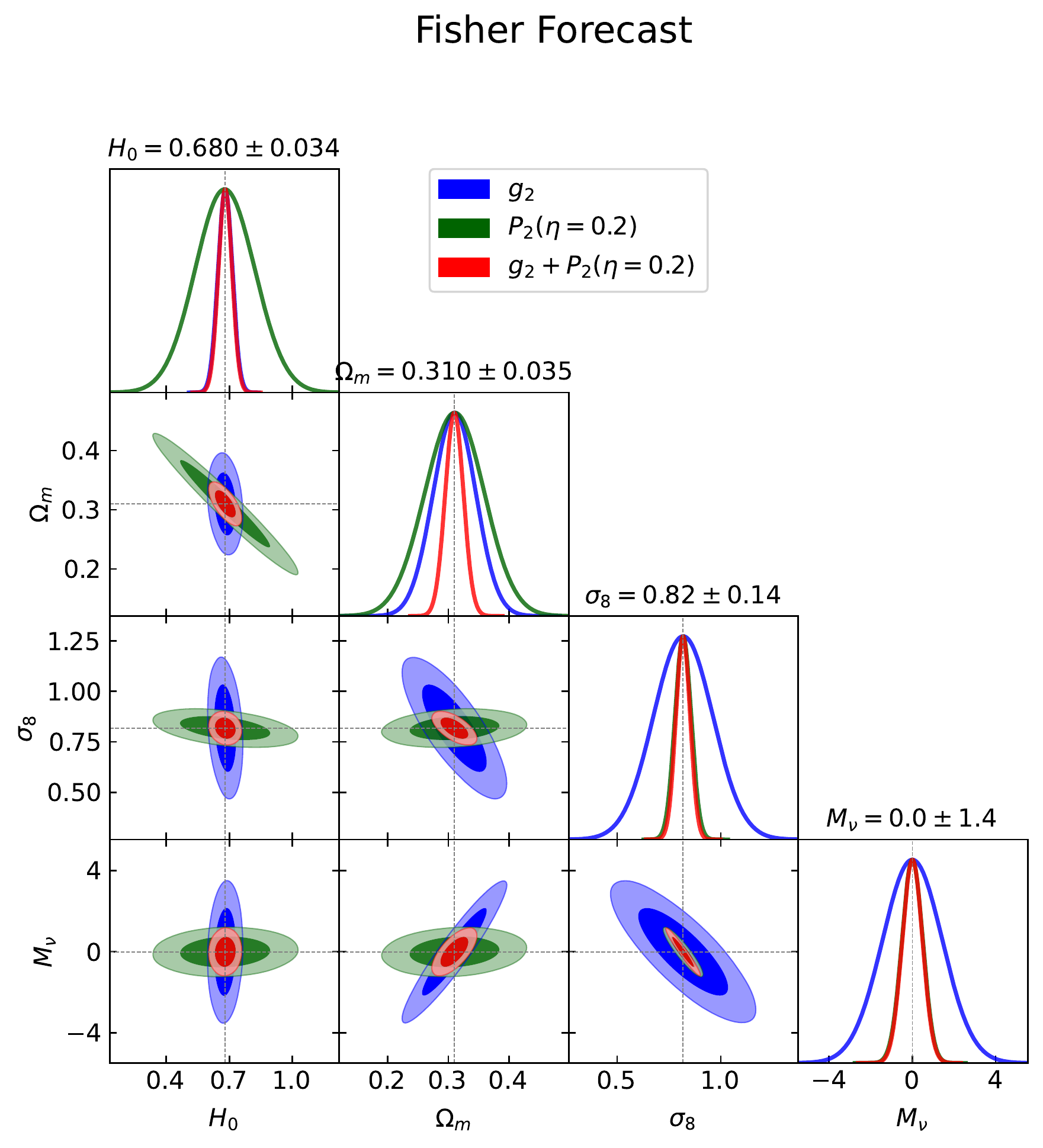}
    \includegraphics[width=0.45\textwidth]{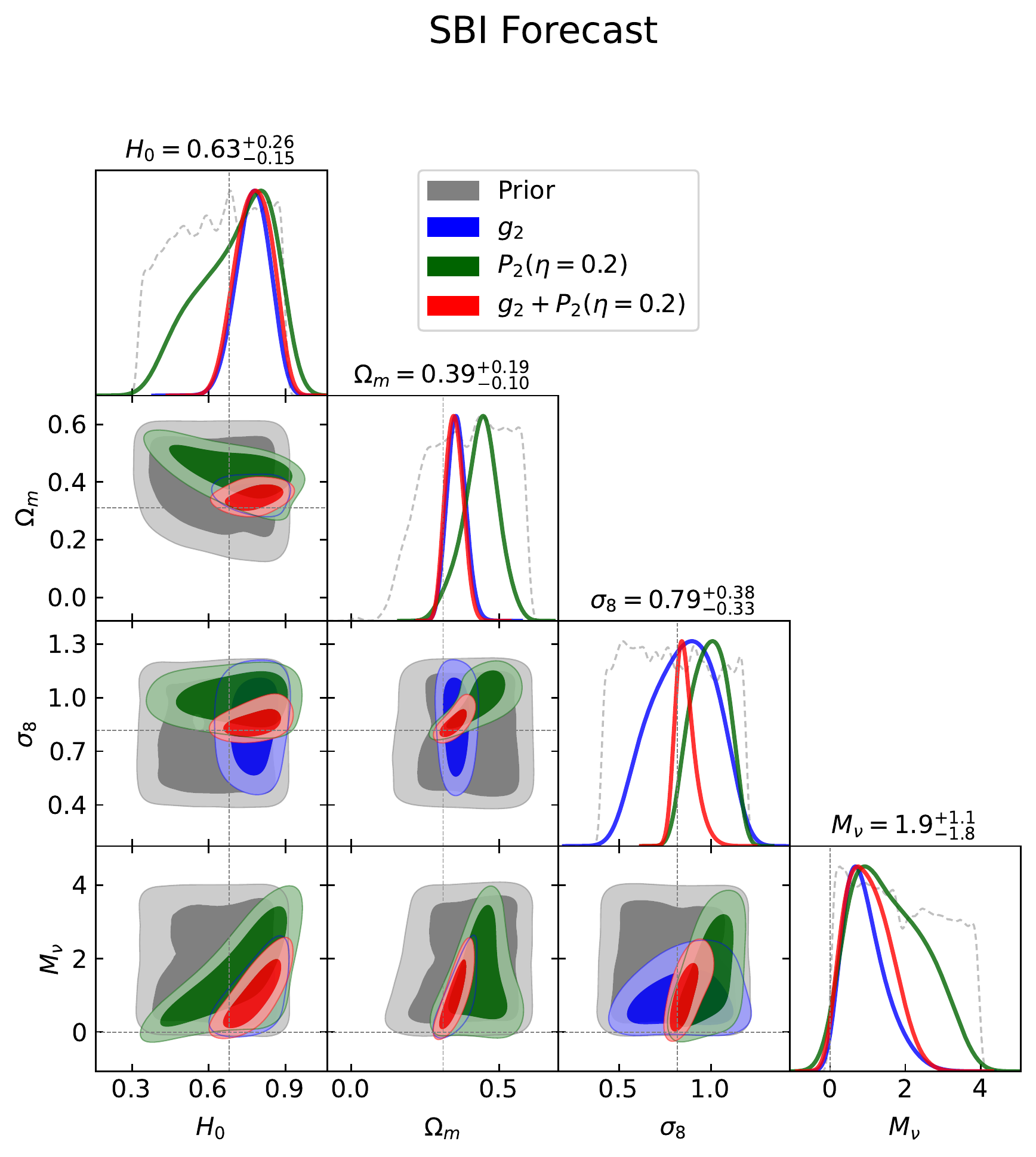}
    \caption{Forecasted constraints on key cosmological parameters using the pair correlation function $g_2$, and the pair-connectedness function, $P_2$, using a Fisher forecast (left) and simulation-based inference (right). The dark (light) ellipses represent our $68\%$ ($95\%$) confidence intervals on the various parameters possible from a single galaxy simulation (as in Fig.\,\ref{fig: density}), with the grey regions show the prior used in the simulation-based approach (from 8192 simulations). We show results for four parameters: $H_0$, giving the Universe's expansion rate, the matter density $\Omega_m$, the clustering amplitude $\sigma_8$ and the total neutrino mass $\sum m_\nu$. The diagonal figures give the marginalized constraints on single parameters, whilst the off-diagonals show the correlations, e.g., the bottom left panel shows the correlation between $h$ and $\sigma_8$. We find that $P_2$ is a poor predictor of $H_0$ and $\Omega_m$ (shown by the broad contours), but can tightly constrain $\sigma_8$. The combination $g_2 + P_2$ significantly increases the precision by which this parameter can be measured, and we find similar results from the Fisher forecasts (which are, in general, overoptimistic), and the simulation-based inference (which is usually conservative).}
    \label{fig: g2-P2-forecast}
\end{figure}

\subsection{Results}
Figure\,\ref{fig: g2-P2-forecast} shows the constraints on cosmological parameters from analyses using the pair correlation function and pair-connectedness function, both via the Fisher and SBI forecasts. From the Fisher forecast, we observe that $P_2$ is a poor predictor of the expansion rate and matter density, but gives much tighter constraints on the clustering amplitude and neutrino mass than $g_2$. This is unsurprising: the slope of $P_2(r)$ is strongly sensitive to the galaxy clustering properties (set by $\sigma_8$, and, on small scales, $\sum m_\nu$), but (being a monotonic function) contains little information on other properties such as early Universe physics. The SBI forecasts give qualitatively similar results, with the $P_2$ constraints on $H_0$ and $\Omega_m$ being largely dominated by the prior, with the combined $g_2+P_2$ constraints reproducing those of $g_2$ alone, For $\sigma_8$, $P_2$ is again shown to be of considerable use, with a significant (factor of a $\approx 2.6\times$, equivalently to observing a seven times greater volume of space) tightening in the one-dimensional posterior found by adding $P_2$, driven by the differing degeneracy directions in the $\sigma_8-H_0$ and $\sigma_8-\Omega_m$ planes. In contrary to the Fisher result, the SBI forecast suggests that the pair-connectedness function does not give significant additional information on the neutrino mass, however, it is shown to change the $\sigma_8-\sum m_\nu$ degeneracy direction considerably.

The disagreement between Fisher and SBI forecasts both quantitatively (in terms of the reduction in width of the $\sigma_8$ posterior) and qualitatively (whether the $\sum m_\nu$ posterior is affected) may appear a little unsettling. We attribute this to a number of reasons: (1) as mentioned above, the Fisher forecast will give artificially narrow constraints if insufficient simulations have been run, (2) the Fisher forecast is inaccurate for parameters whose posterior is non-Gaussian (such as the neutrino mass, due to the $\sum m_\nu>0$ constraint), (3) the SBI forecasts can be artificially \textit{broadened} by insufficient simulations being run. However, the results of Fig.\,\ref{fig: g2-P2-forecast} are enough to convince us that $P_2$ contains significant information regarding the clustering amplitude $\sigma_8$, and its inclusion greatly aids cosmological analyses, including via degeneracy breaking with $\sum m_\nu$. Although we present results only for reduced density $\eta=0.2$ here, a similar story holds also for $\eta=0.1$; in this case, the improvements in cosmological parameters are somewhat weaker, due to the higher-order correlator contributions to $P_2(r,\eta)$ being suppressed \eqref{eq: P2-formal-expan}. \resubtwo{We expect that combining measurements of the pair-connectedness function with multiple values of $\eta$ could further increase the constraining power, again at little computational cost.}

\begin{figure}
    \centering
    \includegraphics[width=0.45\textwidth]{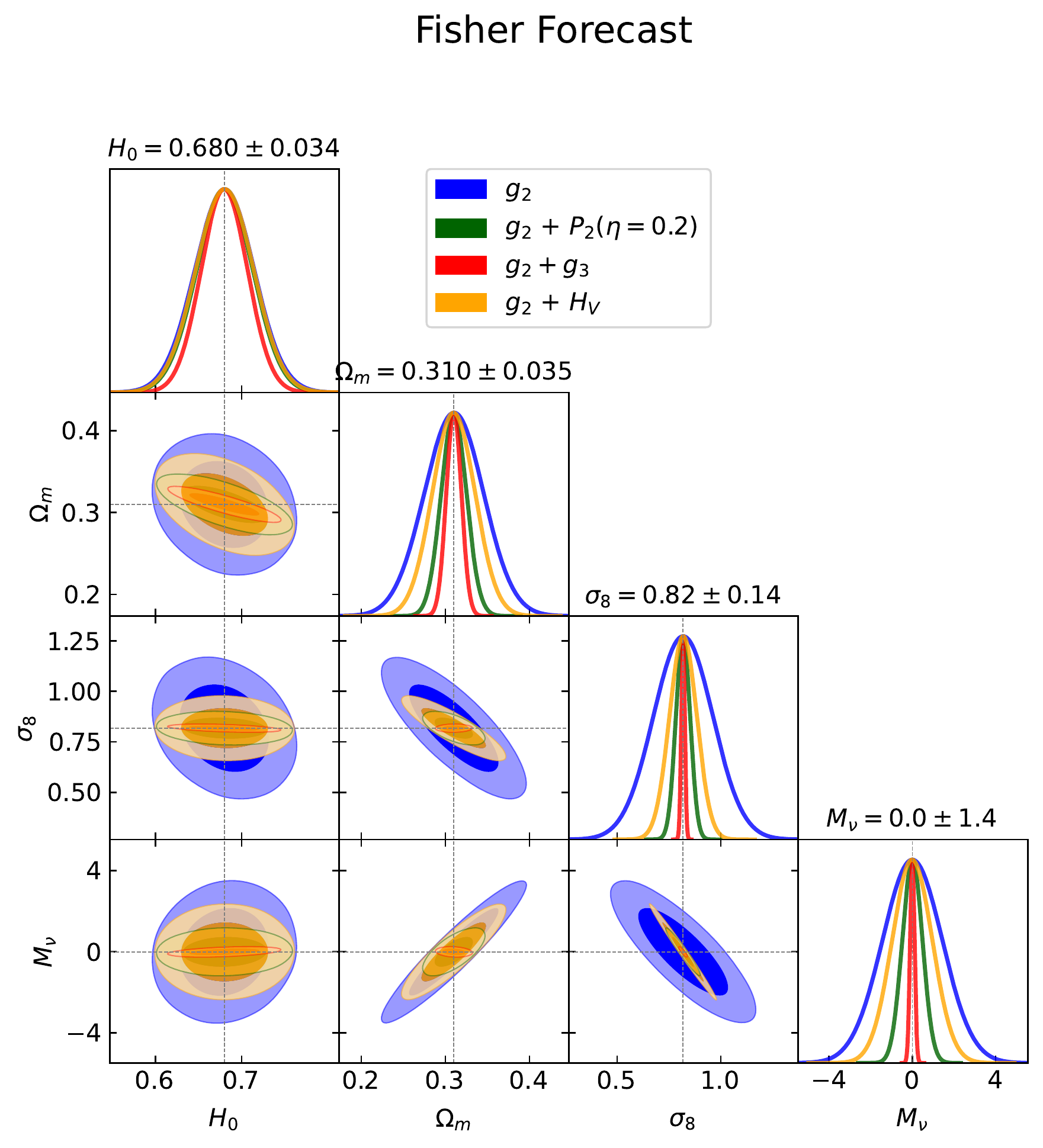}
    \includegraphics[width=0.45\textwidth]{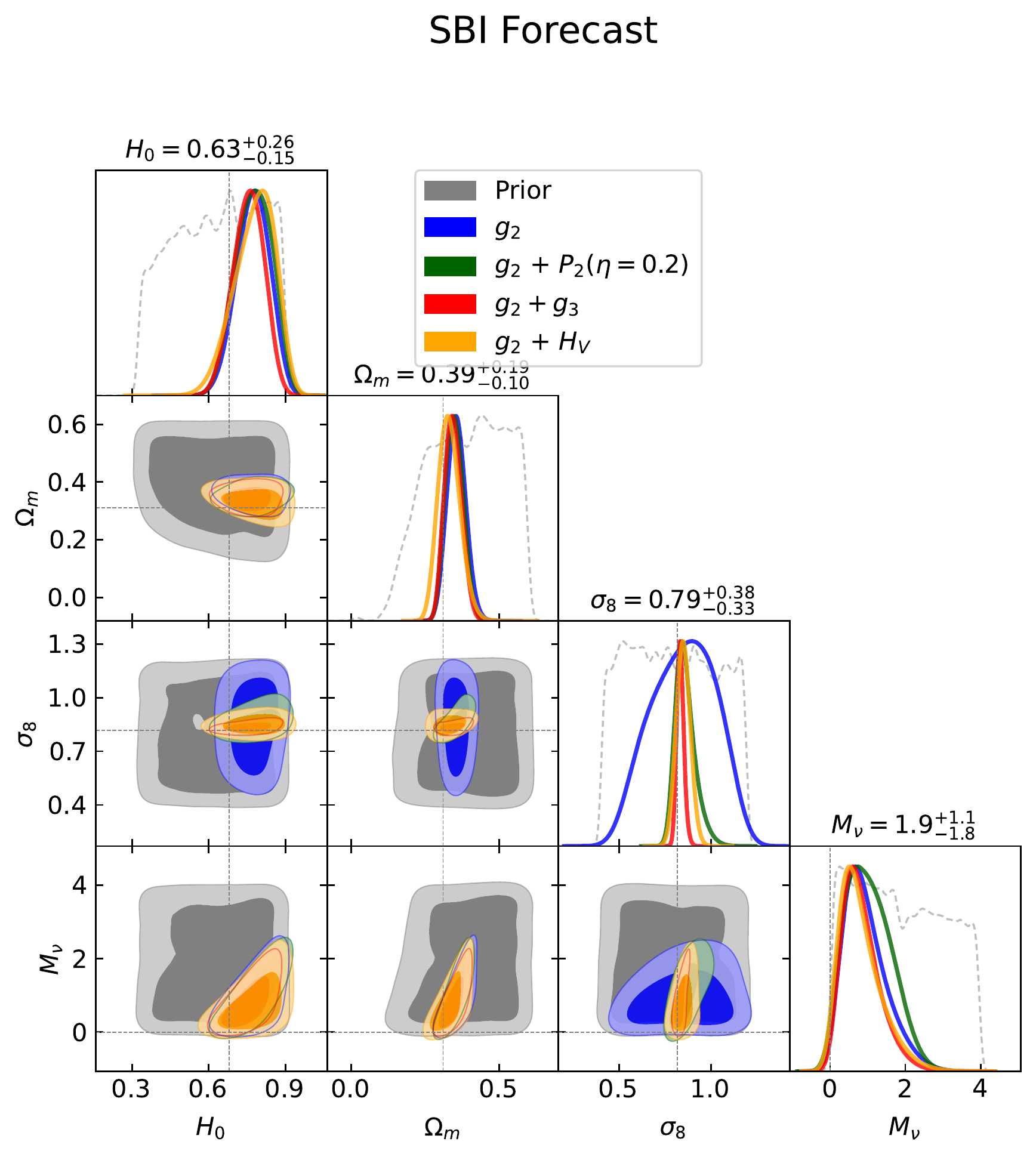}
    \caption{As Fig.\,\ref{fig: g2-P2-forecast} but comparing the constraining power from the combination of the pair correlation function with the pair-connectedness function (green), the three-particle correlation function (red) and the void nearest-neighbor function (yellow). The pair-connectedness function and void nearest-neighbor functions are found to perform almost as well as the three-particle function in this scenario (in terms of constraining $\sigma_8$), and are much faster to compute and analyze.}
    \label{fig: g2-comb-forecast}
\end{figure}

It is interesting to compare the cosmological utility of the pair-connectedness function to that of other higher-order statistics. \resubtwo{Before doing so, let us briefly outline our predictions. In this test, we are limited to relatively large scales ($r\gtrsim 20\Mpch$, due to simulation resolution effects), where the galaxy distribution (if treated as a continuous field) is close to Gaussian. As such, we expect the majority of the information content on cosmological parameters to be encapsulated by the two- and three-point functions, $g_2$ and $g_3$, with only a small amount leaking into higher-order statistics. In this case, the combination of $g_2$ with alternative statistics will likely perform worse than that of $g_2$ and $g_3$; our question is whether there are statistics that are able to recoup most of the information present in $g_3$ in a simpler form (for example in the unidimensional $H_V$ and $P_2$ statistics). If such a statistic exists, it is likely that it also contains significant information on small scales (as probed by future surveys such as that of the Subaru Prime Focus Spectrograph and the DESI Bright Galaxy Survey \citep{DESI:2016fyo,PFSTeam:2012fqu}), where the perturbative hierarchy described above breaks down. Indeed, statistical physics provides examples of small-scale systems where $g_2+P_2$ outperforms $g_2+g_3$ \citep[e.g.,][]{jiao+09}; it will be interesting to study such effects further in the future.}

In Fig.\,\ref{fig: g2-comb-forecast} we show the Fisher and SBI constraints on the same parameter set as above for $g_2$ in combination with $g_3$, $P_2$, and $H_V$.\footnote{\resub{We recall that our $H_V$ void statistic follows a somewhat different definition to the void size function often used in cosmology, and is restricted to comparatively large scales ($r>20\Mpch$, as for the other statistics), with only spherical voids.} This differs from the approach used in several cosmological studies \citep[e.g.,][]{Pisani:2019cvo,Kreisch:2018var,Kreisch:2021xzq} and explains the reduced utility found herein, and the different correlation properties seen in Fig.\,\ref{fig: corr-matrix}.} From both the Fisher and SBI forecasts, we find that no additional statistics lead to significant improvements in the expansion rate constraints, except for a slight tightening from $g_3$. This is not surprising: $H_0$ is primarily measured from an oscillatory feature in $g_2$ arising from acoustic waves in the early Universe, which is generally absent in other statistics. In the SBI forecast, the same is true for the matter density, $\Omega_m$ and the neutrino mass, though the Fisher forecasts disagree on this aspect, as above, and should therefore be taken with a grain of salt (especially given the larger dimensionality of $g_3$). For the clustering amplitude, $\sigma_8$, we find similar improvement when combining $g_2$ with any other statistic, with a slight preference for $g_3$ in the SBI analysis (or a significant one for the Fisher forecast). \resubtwo{This matches the above predictions.}  

Our conclusion from this exercise is the following: if one wishes to constrain the Universe's clustering amplitude (a key target of modern-day cosmology), the addition of $P_2$ or $H_V$ into cosmological analyses provides an excellent route \new{\citep[cf.,][]{2021MNRAS.500.5479B}}, and contains similar information to $g_3$. \resubtwo{Importantly, the alternative statistics are of much lower dimension than $g_3$ and $P_2$ is much less computationally expensive to measure (requiring $\sim 5$ CPU-minutes per simulation, instead of $\sim 1$ CPU-hour for $g_3$ or $H_v$)}.
Furthermore, if performs the analysis using a combination of pair-connectedness functions with different values of $\eta$, the results may be stronger still. Whilst this analysis is necessarily simplistic \resubtwo{and limited to comparatively large scales (due to the nature of the simulation suite)}, 
\resubtwo{it nevertheless} suggests that the pair-connectedness function is a new statistic of significant potential, \resubtwo{and could carry important information also on small scales}. \resub{In contrast to correlators such as $g_2$ and $g_3$ \citep[e.g.,][]{Ivanov:2021kcd}, this is difficult to model analytically even at large $r$, due to its inherent dependence on short-scale physics including the connection between galaxies and dark matter. For this reason it will likely prove useful to adopt a simulation-based methodology to analyze $P_2$, such as the SBI techniques discussed above, and marginalize over parameters controlling galaxy formation.} \resubtwo{On small scales, a similar approach is required for any statistic, due to the breakdown of perturbative modeling.}

\section{Summary}\label{sec: summary}







In this work, we have considered the application of the theory of disordered heterogeneous media and statistical mechanics to cosmology, and of cosmology to the former. By treating the distribution of galaxies in the present-day Universe as a point process, we can analyze the data using techniques developed to characterize heterogeneous media, such as the correlation functions and nearest-neighbor distributions. Furthermore, by augmenting the data-set with some concept of `connectedness' (here defined by circumscribing the galaxies with spheres), we may utilize various clustering diagnostics and pair-connectedness functions, which encode a different subset of the information present within the distribution and additionally allows the percolation properties to be determined. Such a framework (a) provides a novel method for understanding the galaxy distribution, whose importance will only grow in the next decade with the plurality of upcoming telescopes, and (b) demonstrates the applicability of heterogeneous media and statistical mechanical techniques in a very different regime to that usually explored.

Our main conclusions are the following: 
\begin{itemize}
    \item The galaxy distribution exhibits very different physical properties to those of conventional materials, leading to distinct signatures in a wide variety of clustering and correlation descriptors. On the largest scales, the system approaches hyperuniformity, whilst on the the smallest, it becomes almost antihyperuniform and strongly inhomogeneous; this dichotomy arises from the fact that the minimum separation between galaxies is much smaller than the mean interparticle distance, with localized groups of galaxies separated by vast cosmic distance.
    \item Physically, the cosmological system has two peculiarities: (a) although we treat the galaxies as point objects, they have some physical scale in practice, and cannot overlap, (b) the distribution carries the signatures of a large-scale stochastic background that modulates the quasi-Poissonian distribution; this is sourced by early Universe physics and gravitational evolution,
    \item The galaxy pair correlation function shows this behavior clearly, with the expected hyperuniform tail appearing only at gargantuan scales ($r\gtrsim 200\Mpch$), and with a sharp peak at the \resub{mean pairwise particle} separation of $r\approx 8\Mpch$. This scale separation induces a large number variance, which is unusually super-Poissonian on small scales, yet sub-Poissonian on the largest. These results are consistent with the order metric, $\tau$, which we determine for the galaxy sample for the first time: its value ($\tau=4.85$) implies that the system is strongly correlated and disordered. The nearest-neighbor functions are again consistent with this picture, with considerably extended tails, a lack of particle pairs below some critical galaxy size, and much enhanced variance relative to the Poissonian case (and most other common scenarios).
    \item Analysis of pair-connectedness functions, mean particle numbers, and sample spanning clusters indicate that the galaxy sample percolates at significantly lower reduced densities than corresponding Poisson realizations. For the fiducial galaxy simulations, finite-scaling analysis gives $\eta_c = 0.25$ in the former case compared to $\eta_c = 0.34$ in the latter, a difference which is amplified by increasing the sample density.  This is again supported by the above evidence: the scale separation is a consequence of the extra small-scale clustering in the galaxy distribution which leads to faster percolation. Both scenarios appear to have the same critical exponents and fractal dimensions, implying that they live in the same universality class, despite very different physics operating.
    \item The pair-connectedness function \resubtwo{is a conceptually straightforward and easy-to-measure statistic that} carries useful and accessible large-scale information about the underlying physical parameters of the Universe, \resubtwo{and can be trivially extended to small scales}. \resubtwo{This} could enhance the cosmological utility of future galaxy surveys, \resubtwo{in combination with conventional techniques}. Using simulation-based analysis techniques, we forecast that constraints on amplitude of clustering improve by a factor of $\approx 5$ (or $\approx 25$ in terms of survey volume) when performing inference using the large-scale pair-connectedness and pair correlation functions as opposed to the pair correlation function alone (which is standard in cosmology). This provides a useful alternative to the three-particle correlation function $g_3$, which is of significantly lower dimension and much faster to model, and is shown to be a resummation of correlation functions of all order. Unlike the large-scale three-particle function, it seems unlikely that $P_2(r)$ can be modelled analytically; simulation-based treatments will likely be required in this case.
\end{itemize}

The galaxy samples used in this work are purposefully simplified, in order to provide a proof-of-concept study capturing the essential physical attributes of the cosmological set-up. More work is needed before the statistics can be applied to real data, and will require the following: \resubtwo{(a) higher resolution simulations containing more particles, allowing smaller scales to be probed,} (b) inclusion of real galaxies in the simulations, rather than dark matter halos, and the associated physical uncertainties with their formation \citep[e.g.,][]{Desjacques:2016bnm,wechsler18}, (c) anisotropic distortions in the Universe created by transforming from redshifts to physical coordinates \citep{Kaiser:1987qv}, (d) inhomogeneities in the field induced by observational effects, such as the limited field-of-view of the telescope. However, all of these complexities have been overcome a number of times before for other statistics (such as the correlation functions \citep[e.g.,][]{philcox22}) and we expect can be similarly surmounted in this case. \new{Furthermore, it is important to characterize how the statistics depend on the galaxy sample: for the correlation functions, this is well understood (and encapsulated by `bias parameters', which depend on galaxy mass and luminosity), but should be explored further for nearest-neighbor and pair-connectedness functions, as well as the percolation threshold.}

Finally, we consider the broader extensions of this work. Although we have restricted our gaze to galaxy distributions, this is far from being the only stochastic distribution in the Universe. One additional application could include a more principled treatment of cosmic voids \citep[e.g.,][]{Pisani:2019cvo,Sheth:2003py}: these are low-density regions in the galaxy distribution that form a partition of the space, and could be described by the same mathematics as that invoked for percolation. Even more relevant is the distribution of `bubbles' of ionized gas around the first galaxies \citep[e.g.,][]{loeb01,Lee:2007dt}. The growth of such bubbles likely led to the Universe's reionization approximately one billion years after the Big Bang, the time of which is set by percolation itself. Finally, we note that there are a wealth of  techniques from the theory of disordered heterogeneous media that have not been considered in this work. It would be interesting to consider the utility of the various descriptors in the context of `simulated annealing' \citep[e.g.,][]{Ye98a,jiao+09}, to understand the extent to which any statistic can capture the full complexities of the field, though we caution that conventional approaches will likely need to be modified to account for the peculiarities of the galaxy distribution, in particular its significant scale separation. Further still, we may consider how annealing techniques allow us to recover `effective pair interactions' between individual galaxies \citep{To22d}, and thus learn more about the Universe's average dynamics. 

\begin{acknowledgments}
\footnotesize
\new{We thank Jim Peebles, Paul Steinhardt, Robert Scherrer, Arka Banerjee, Tom Abel, and \resub{Haina Wang} for insightful comments on this manuscript.} OHEP is additionally grateful to Alice Pisani and Will Coulton for useful discussions regarding cosmic voids and Fisher forecasts respectively. \resubtwo{We additionally thank the anonymous referees for an insightful report. OHEP is a Junior Fellow of the Simons Society of Fellows} and thanks the Institute for Advanced Study for their hospitality and abundance of baked goods.  ST thanks the Institute for Advanced Study for their hospitality during his sabbatical leave there.

\vskip 4 pt

The authors are pleased to acknowledge that the work reported in this paper was substantially performed using the Princeton Research Computing resources at Princeton University, which is a consortium of groups led by the Princeton Institute for Computational Science and Engineering (PICSciE) and the Office of Information Technology's Research Computing Division. Additional computations were performed on the Helios cluster at the Institute for Advanced Study, Princeton.
\end{acknowledgments}

\appendix 


\providecommand{\href}[2]{#2}
\begingroup\raggedright\endgroup

\end{document}